\documentclass[10pt]{buildsys-proc}

\usepackage{balance}
\usepackage{comment}
\usepackage[dvipsnames]{xcolor} % Defined by Manoj
\usepackage{textcomp} % Defined by Manoj used in single quotes
\usepackage{graphicx} % Defined by Manoj
\usepackage{caption} % Defined by Manoj
\usepackage{subcaption} % Defined by Manoj
\usepackage{epstopdf}
\usepackage{epsfig}
\usepackage{url}
\usepackage[bottom,flushmargin,hang,multiple]{footmisc} % Manoj Used this to add foot note
\usepackage{color}

\numberofauthors{3}

\author{
\alignauthor Manoj Gulati\\
       \affaddr{IIIT, Delhi}\\
       \email{manojg@iiitd.ac.in}
\alignauthor Shobha Sundar Ram\\
       \affaddr{IIIT, Delhi}\\
       \email{shobha@iiitd.ac.in}
\alignauthor Amarjeet Singh\\
       \affaddr{IIIT, Delhi}\\
       \email{amarjeet@iiitd.ac.in}
}

\title{An In Depth Study into Using EMI Signatures for Appliance Identification}

\crdata{978-1-4503-3144-9}
\conferenceinfo{BuildSys'14,} {November 5--6, 2014, Memphis, TN, USA.}
\CopyrightYear{2014}

\begin{document}

\maketitle
\raggedbottom

\begin{abstract}
Energy conservation is a key factor towards long term energy sustainability. Real-time end user energy feedback, using disaggregated electric load composition, can play a pivotal role in motivating consumers towards energy conservation. Recent works have explored using high frequency conducted electromagnetic interference (EMI) on power lines as a single point sensing parameter for monitoring common home appliances. However, key questions regarding the reliability and feasibility of using EMI signatures for non-intrusive load monitoring over multiple appliances across different sensing paradigms remain unanswered. This work presents some of the key challenges towards using EMI as a unique and time invariant feature for load disaggregation. In-depth empirical evaluations of a large number of appliances in different sensing configurations are carried out, in both laboratory and real world settings. Insights into the effects of external parameters such as line impedance, background noise and appliance coupling on the EMI behavior of an appliance are realized through simulations and measurements. A generic approach for simulating the EMI behavior of an appliance that can then be used to do a detailed analysis of real world phenomenology is presented. The simulation approach is validated with EMI data from a router. Our EMI dataset - High Frequency EMI Dataset (HFED) is also released. 
\end{abstract}

% \category{H.4}{Information Systems Applications}{Miscellaneous}
% A category including the fourth, optional field follows...

\category{D.2.8}{Power monitoring, load disaggregation}{Metrics}[measurability]

\terms{SMPS, Electromagnetic Interference (EMI), Alternating Current (AC), Direct Current (DC)}
%\keywords{EMI} % NOT required for Proceedings

\section{Introduction}
  \label{sec:intro}

Buildings across the globe are among the primary consumers of energy (29\% in USA, 24\% in UK and 47\% in India) \cite{halverson2009country, TechnicalreportUK, evans2009country}. With the rapid rate of construction and introduction of new home appliances, energy consumption in buildings is likely to increase in years to come. Due to high transmission and distribution losses, especially in developing countries like India, it is widely accepted that ``1 unit of electricity saved at the consumer end is equivalent to 2 units of electricity produced at the power plant'' \cite{keshav2011internet}. For residential consumers, energy monitoring and disaggregation for each occupant and for each appliance helps in improved understanding of consumption practices that can subsequently, drive 5-15\% reduction in energy consumption \cite{darby2006effectiveness, Survey:Mohit}.

Systems for energy disaggregation can be broadly classified, based on the number of sensing locations, into two classes - distributed direct sensing and single point sensing. Distributed direct sensing, requiring measurement at each appliance, is prohibitively complex from both the deployment and the maintenance perspective. Non-intrusive load monitoring (NILM) is a technique used to disaggregate the electric load composition of a household using single point sensing usually at the mains power feed. Though NILM was originally introduced more than two decades ago \cite{hart1992nonintrusive}, recent large-scale deployments of smart meters by electrical utilities across the world have resulted in increased research interest \cite{farhangi2010path}. However, due to limited data collection capabilities of smart meters being deployed, significant NILM research in the recent past are driven by low frequency data (1 Hz or lower). While there is some work involving NILM using higher frequency (few kHz - MHz) power consumption data, much of it is limited to controlled laboratory experiments \cite{leeb1995transient}. Recently, Gupta et al \cite{gupta2010electrisense} proposed indirectly identifying disaggregated energy consumption using high frequency conducted electromagnetic interference (EMI) that emanate from electronic appliances. They showed that EMI propagates through the power infrastructure and hence can be measured from a single point installation at the home level. A follow up work \cite{enev2011televisions} showed that EMI signals can be further used to get detailed information regarding the operational state of an appliance e.g. the genre of programs being watched on a television.

The main objective of our work is to present some of the key challenges towards exploiting EMI as a unique and deterministic signature for appliance disaggregation. While prior work in power line EMI sensing for NILM may sound promising, much of this work is based upon certain assumptions derived from limited set of experiments. For instance, Gupta et al. \cite{gupta2010electrisense} observed that SMPS based appliances conduct unique and observable EMI. However, their study was restricted to a small set of appliances. Furthermore, they stated that appliances with EMI filters will also conduct observable EMI but the actual impact of such filters on the EMI signature of the appliance and other devices on the power line was not well studied. Miro and Froehlich et al in their work \cite{enev2011televisions, froehlich2011disaggregated}, limited to televisions, observed that EMI signatures remain time-invariant and robust to background noise. Through an extensive set of experiments performed both in lab setting and in a real home, on a wide range of appliances, we show some of the limitations of the above mentioned assumptions: only some SMPS based appliances generate measurable EMI; coherent coupling of EMI from different appliances on a power line may impact the measurability of EMI from each individual appliance; inbuilt EMI filters impact the conducted EMI of the appliance connected to the filter as well as neighbouring appliances on the power line; and time-varying and non-deterministic behavior of background noise on the power line may significantly impact the EMI signatures of an appliance. We further highlight several observations from our measured data that demonstrate the need for better understanding of conducted EMI before making a case for using it reliably for appliance disaggregation. We believe, ours is the first work that brings forth a detailed analysis of some of the challenges of using conducted EMI for appliance disaggregation. We release our dataset- High Frequency Energy Dataset (HFED)\footnote{\url{http://hfed.github.io/}} of EMI traces collected from both the lab and the residential settings. This is the first such dataset released from a developing country wherein the quality of electrical infrastructure and appliances used may have impacted the observable EMI. Finally, we augment our empirical analysis of conducted EMI with basic simulation models to better understand the behavior of conducted EMI and the impact of external factors e.g. line impedance and coupling among appliances.

Our paper is organized as follows. In Section 2, we provide a literature study of different sensing techniques for NILM. Conducted EMI and its specific use for NILM are presented in Section 3. In Sections 4 and 5, we present the experimental setup for gathering conducted EMI data in labs and homes as well as the detailed analysis of the measurement data. We present a generic simulation model that can be fine-tuned to match the EMI from specific appliances in Section 6. This simulation model is used to understand the real-world phenomena observed in the measurement data. We discuss the details of our released dataset in Section 7. Finally, we conclude in Section 8.

\section{Related Work}
    \label{sec:relatedwork}

Much of the NILM work is focused on measuring electrical parameters such as voltage, power factor, active power and reactive power at the mains power inlet  in order to identify the appliances contributing to the overall energy consumption \cite{zoha2012non}. Such a measurement system could be constrained due to complex circuitry and limited space available in the mains panel. On the other hand, a system based on single point EMI sensing could ideally be placed at any location in the home thus making it relatively more convenient to deploy.

Apart from NILM methods utilizing electrical signatures, several indirect techniques have been explored, in the past, for load decomposition \cite{kim2010granger, kim2009viridiscope, rowe2010contactless, srinivasan2013fixturefinder, taysi2010tinyears}. These include monitoring ambient sound levels to detect sound generated by appliances \cite{taysi2010tinyears}; detecting light events based on the variation in light intensity \cite{kim2009viridiscope}; measuring stray magnetic fields generated by certain appliances \cite{rowe2010contactless}; and using combined IP/Wi-Fi traffic activity data with smart meter data for estimating load decomposition \cite{kim2010granger}. However, these approaches have not been widely adopted due to the complexities of integrating a diverse set of sensing modalities across multiple locations in home settings.

More recently, there have been extensive research efforts towards developing modelling techniques to infer constituent loads from power consumption data at the main meter level \cite{parson2012non, zoha2013low, batra2013indic, leeb1995transient, patel2007flick, NILMTK}. Some of these modelling approaches include using the hidden Markov model (HMM) \cite{parson2012non} and its variants such as factorial hidden Markov model (FHMM)\cite{zoha2013low}, combinatorial optimization \cite{batra2013indic} and transient based analysis \cite{leeb1995transient, patel2007flick}. Some of these modelling approaches could be extended to EMI based appliance disaggregation. 
Many of the existing NILM techniques, using low frequency signatures, either under-perform or fail due to time-varying power consumption patterns of these appliances \cite{barker2013empirical, dong2014fundamental}. NILM using high frequency sensing is a field that is relatively less explored. Patel et al. used low amplitude transients (up to a few kHz) for detecting electrical events \cite{patel2007flick, lee2004exploration}, and showed that these events, combined with smart meter data, can help in appliance disaggregation. Leeb et al. used low frequency harmonics up to a few kHz to disaggregate appliances and diagnose faults in electrical systems \cite{orji2010fault, shaw2008nonintrusive}. Other studies have utilized low frequency power signatures, either harmonics of 50 Hz sinusoid or transient noise generated by appliance switching, to distinguish household appliances \cite{lee2004exploration, patel2007flick}. Recent works have used high frequency EMI, generated from high-speed switching circuits to identify certain home appliances \cite{enev2011televisions, froehlich2011disaggregated, gupta2010electrisense}. These switching circuits form a fundamental component of switched mode power supplies (SMPS) found in most electronic appliances today. 

Our work aims to establish better understanding of the challenges of using high frequency EMI signatures for NILM. Several of the insights discussed in this work either refute claims made in prior work or bring forth fresh perspectives based on empirical observations and simulation studies. 

\begin{figure*}[t!]
\centering
  \begin{subfigure}[t]{0.63\textwidth}
    \includegraphics[width=\textwidth]{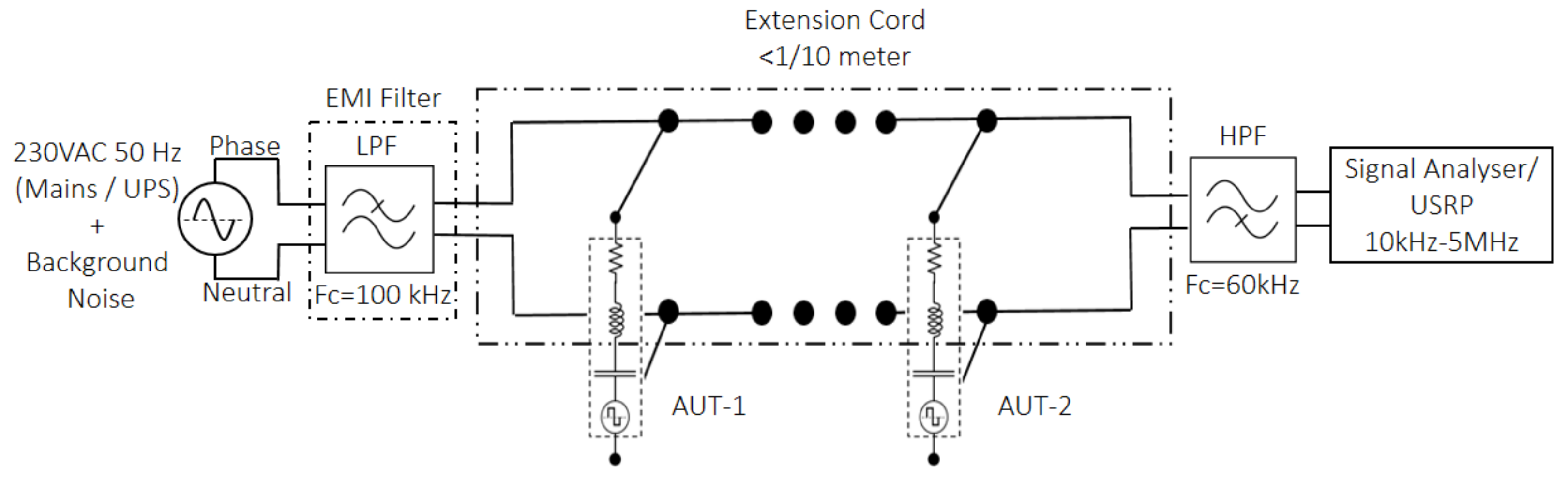}
      \caption{EMI Measurement Setup}
%   \label{fig:Instance11}
    \label{fig:Testsetupslab}
  \end{subfigure}
  \begin{subfigure}[t]{0.34\textwidth}
      \includegraphics[width=\textwidth]{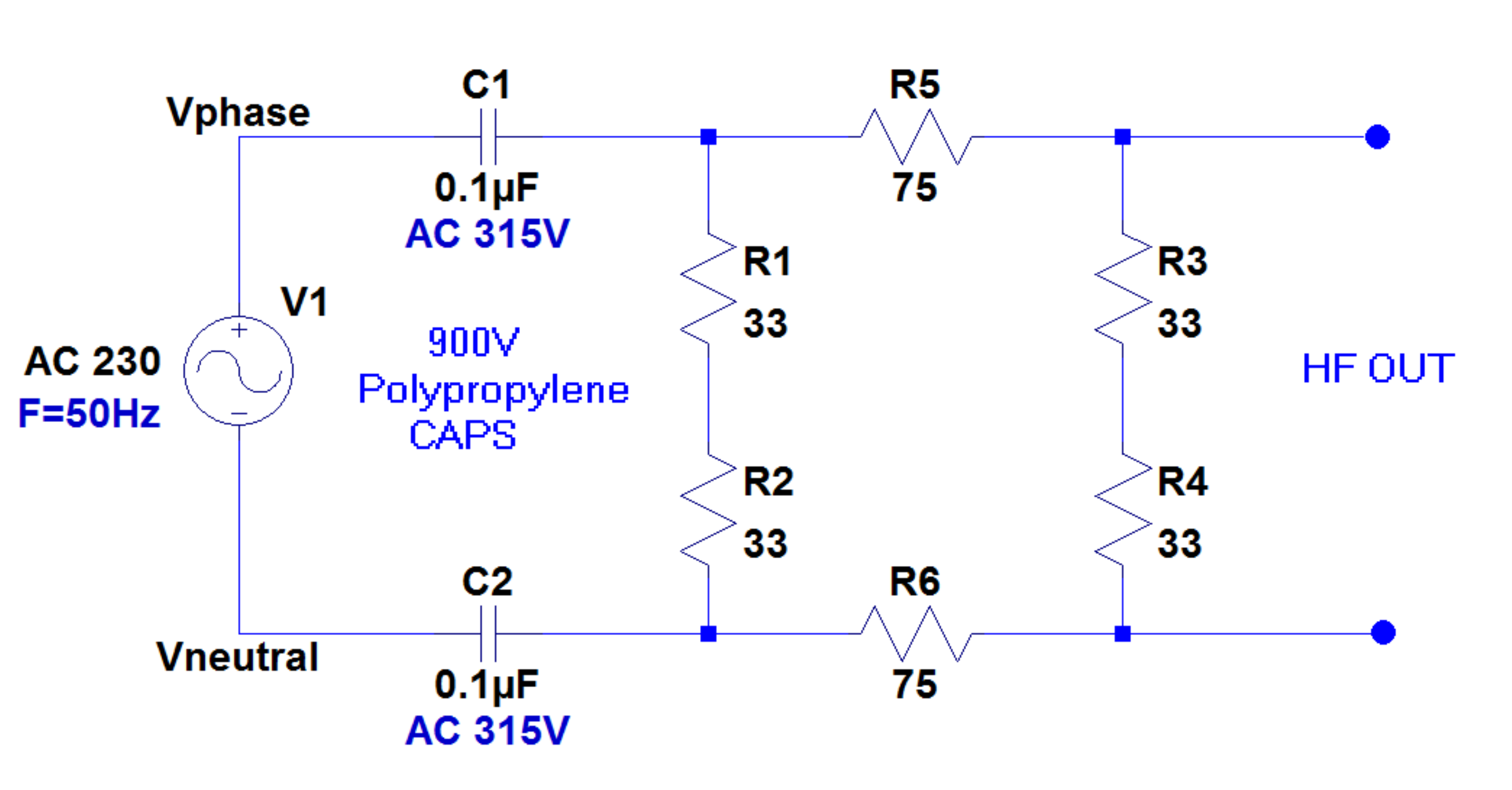}
        \caption{Differential HPF schematic}
  %     \label{fig:Instance31}
        \label{fig:schematic}
  \end{subfigure}
  \vspace{-3mm}
  \caption{Test setup used for EMI measurements. Appliances under test (AUTs) were connected through an optional low pass filter (LPF) to the mains power line. Observations after filtering through a differential high pass filter (HPF), shown in detail in part(b), were made either with a signal analyser or a universal software radio peripheral(USRP).}
    \label{fig:setupDetails}  
    \vspace{-3mm}      
\end{figure*} 

\section{Insights into Conducted EMI}
    \label{sec:insights}

EMI is wideband (9 kHz- 30 MHz) noise, generated by electronic appliances, that is conducted either through power lines or radiated into the ambient environment. We focus only on conducted EMI in this work. Conducted EMI from many home appliances, such as LCD TVs, laptop chargers, mobile chargers, modems and CFLs, arise mostly from SMPS. SMPS are popular compared to linear regulator based power supplies because of their high efficiency, small size and low cost. High speed switching circuits draw non-linear current from the power supply, giving rise to high frequency harmonics of their switching frequencies \cite{paul2006introduction}. EMI is, therefore, an unwanted signal that may interfere with the functioning of other electronic appliances, connected to the same power line and may degrade their performances. Several governing bodies, across the globe, have laid down regulations quantifying the maximum amount of EMI that a particular appliance can introduce into the power line. As per IEC\footnote{International Electrotechnical Commission}, the frequency range of conducted EMI lies from 9 kHz to 30 MHz. Some of the standards governing the upper bounds for conducted EMI are CISPR 11/EN 55011, CISPR 14-1/EN 55014-1 and FCC15/18 for industrial and household products and CISPR 13/EN 55013 for consumer products and radio receivers. 

Each SMPS has a unique and constant switching frequency based upon the power requirements of the load appliance \cite{boost2007switch}. Due to this unique characteristic, prior work explored if EMI signals are time-invariant feature vectors that could be leveraged for disaggregating appliances. Despite the standards and limits imposed on conducted EMI by regulating authorities, previous studies showed that even reduced levels of EMI (less than -40dBm) from some appliances were observable with measuring instruments of high sensitivity and could be exploited for monitoring loads. In order to effectively use EMI signals for appliance disaggregation, a thorough understanding of several of the following characteristics is required:
\begin{itemize}
\item The impedance of a power line attenuates the conducted EMI. Do the EMI characteristics of an appliance vary when observed at different places within a home?
\item Do conducted EMI from multiple appliances interfere with each other? Are the interference mechanisms unique to an appliance or will they vary depending on the other appliances that are connected to the power line? 
\item Does background noise from the power line infrastructure that may differ from one home to another, significantly impact the observed EMI from an appliance?
\item Do all appliances follow specified EMI standards? 
\item Do all appliances of the same category, e.g. CFLs from different manufacturers, exhibit similar EMI characteristics? 
\end{itemize}

In this work, we aim to take a step forward in answering many of the above concerns through an extensive empirical study performed on a set of 24 loads and 4 test setups across lab and residential settings.  

\begin{table*}[!ht]
\begin{tabular}{|l|l|p{2.2in}|l|p{0.8in}|p{0.7in}|}
\hline
\multicolumn{1}{|c|}{S.No.} & \multicolumn{1}{c|}{AUTs} & \multicolumn{1}{c|}{Brand}                                       & \multicolumn{1}{c|}{Category} & Power Ratings (in Watts) & Location Used \\ \hline
1                           & CFL1, 2, 3, 4            & Crompton Greaves {[}1{]}, Bajaj {[}2, 3, 4{]}                    & SMPS                          & 18, 15, 15, 5            & L, R          \\
2                           & LED Lamp-1, 2, 3          & Genre India {[}1{]}, Unbranded {[}2{]}, Crompton Greaves {[}3{]} & SMPS                          & 5, 3, 0.5               & L, R          \\
3                           & Laptop Charger-1, 2       & Dell {[}1{]}, HP {[}2{]}                                         & SMPS                          & 90, 65                   & L, R          \\
4                           & Phone Charger-1, 2, 3     & Samsung {[}1{]}, Asus {[}2{]}, LG {[}3{]}                        & SMPS                          & 5, 7, 6                 & L, R          \\
5                           & LCD Monitor               & HP P191                                                         & SMPS                          & 20                       & L, R          \\
6                           & Printer                   & HP P1007                                                              & SMPS                          & 5                        & L, R          \\
7                           & Speakers                  & Harman Kardon                                                    & SMPS                          & 24                       & L, R          \\
8                           & Modem                     & Asus Router                                                      & SMPS                          & 18                       & L, R          \\
9                           & Induction Cooktop -1, 2   & Philips {[}1{]}, Maharaja Whiteline {[}2{]}                      & SMPS                          & {[}500,1300{]}, {[}600,1000{]}     & L, R          \\
10                          & Microwave                 & Kenstar                                                          & -                             & 1250                    & R             \\
11                          & Refrigerator              & LG                                                               & NON SMPS                      & 1020                    & R             \\
12                          & Blender                   & Inalsa                                                           & NON SMPS                      & 180                     & L, R          \\
13                          & Iron                      & Philips                                                          & NON SMPS                      & 535                     & L             \\ 
14                          & Room Heater               & North Star                                                       & NON SMPS                      & 1500                    & L             \\ 
15                          & Television                & LG                                                               & SMPS                          & 60                      & R             \\ \hline
\end{tabular}
\caption{List of appliances for which the EMI measurements were done in the lab (L) and the residential (R) settings. Same class of appliances are placed together in a row  and the corresponding items in subsequent columns are comma separated to specify the respective order. Induction cooktop measurements have two specified power ratings each representing different operational modes.} 
\label{tab:listofappliances}
\end{table*}

\section{Experimental Setup}
    \label{sec:setup}
    
Figure-\ref{fig:setupDetails} shows details of the experimental setup used for collecting EMI data from the 230V power line, for multiple appliances, in both the laboratory and the residential settings. The mains power line consists of 50 Hz power signal, background noise from the electrical infrastructure of the building (such as air handling units, variable speed drives and lighting sources) and EMI introduced by the appliances under test (AUTs).  To isolate the background noise for some experiments, the mains power supply is replaced with Luminous 600VA uninterrupted power supply (UPS). AUTs are connected to the mains either through an off-the-shelf power extension cord, having a distance of few centimeters between plug points or, a custom built 10 meter long extension cord, with a plug point at every 2 meters. %Both of these extension cords are connected to the mains power supply. 
Some off-the-shelf extension cords\footnote{Belkin BE112230-08} come with built-in low-pass EMI filters, to isolate the appliances from noise in the power supply. To analyze the impact of such filters, an Elcom EP-15AP power line EMI filter offering an attenuation of approximately 10 to 20 dB from 1 MHz to 100 MHz is used for some of the experiments. 

Though IEC specifies the range of conducted EMI from 9 kHz to 30 MHz, the AUTs , in our measurements, show EMI only up to 5 MHz.  A custom built differential high pass filter (HPF), with a cutoff frequency of 60 kHz, is used to pass this high frequency EMI noise to the measurement equipment. The high cutoff frequency is chosen in order to prevent the 50 Hz power signal and its harmonics from damaging the sensitive analog front end of the measurement devices. The detailed schematic of the high pass filter is shown in Figure-\ref{fig:schematic}. While the circuit is similar to the one previously used in \cite{gupta2010electrisense}, the component values are slightly modified for 230V operation. EMI is measured using either the Agilent N9000A CXA signal analyser or a Universal Software Radio Peripheral (USRP) N200.

The high sensitivity and ease of configurability of the signal analyzer, makes it useful for measurements in laboratory settings. On the other hand, portability of USRP makes it suitable for conducting EMI measurements in laboratory, as well as residential settings.\\
EMI measurements in the lab settings are carried out with four different configurations of the experimental setup. In setup-1, an AUT is connected to a nearby plug point (NPP) on the off-the-shelf extension cord. Therefore, the NPP, is a few centimeters from the sensing plug point (SPP) where the signal analyzer is connected. In setup-2, the AUT is connected to a distant plug point (DPP), 10 meters from the SPP, on the custom extension cord. This test is performed to study the impact of power line impedance on the EMI measurements. In both the cases described above, the low-pass EMI filter is not connected to the experimental setup. In setup-3, the impact of a power line EMI filter is studied by incorporating a low-pass filter to setup-1. In these three cases, extension cords are connected directly to the AC mains power supply. In setup-4, a UPS having 3 power outlets, few centimeters apart, is used to power the AUTs. Since the UPS is independently powered, it is isolated from the noise in the building's electrical infrastructure. Due to the limited power capacity of the UPS, EMI measurements with setup-4 could only be carried out for medium and the low power AUTs.

In all four cases, EMI measurements are carried out with both individual AUT or a combination of AUTs, from 10 kHz to 5 MHz. Additionally, background noise on the power line is measured for each case with the AUTs disconnected from the extension cord.
Setup-1 described above is repeated in residential settings while replacing the signal analyzer with the USRP. 
Table~\ref{tab:listofappliances} shows the list of 24 AUTs used in lab and residential settings along with their manufacturer details and power ratings. 

\section{Observations and Analysis}
    \label{sec:observations}
    
We now discuss the impact of a variety of sensing parameters on the EMI signatures of the AUTs, based on observations from the experiments.
% Observation-1 : Not all appliances conduct EMI
\begin{figure*}[t!]
\vspace{-10pt}
\hspace{-7pt}
    \begin{subfigure}{0.6\columnwidth}
    \includegraphics[width=\textwidth]{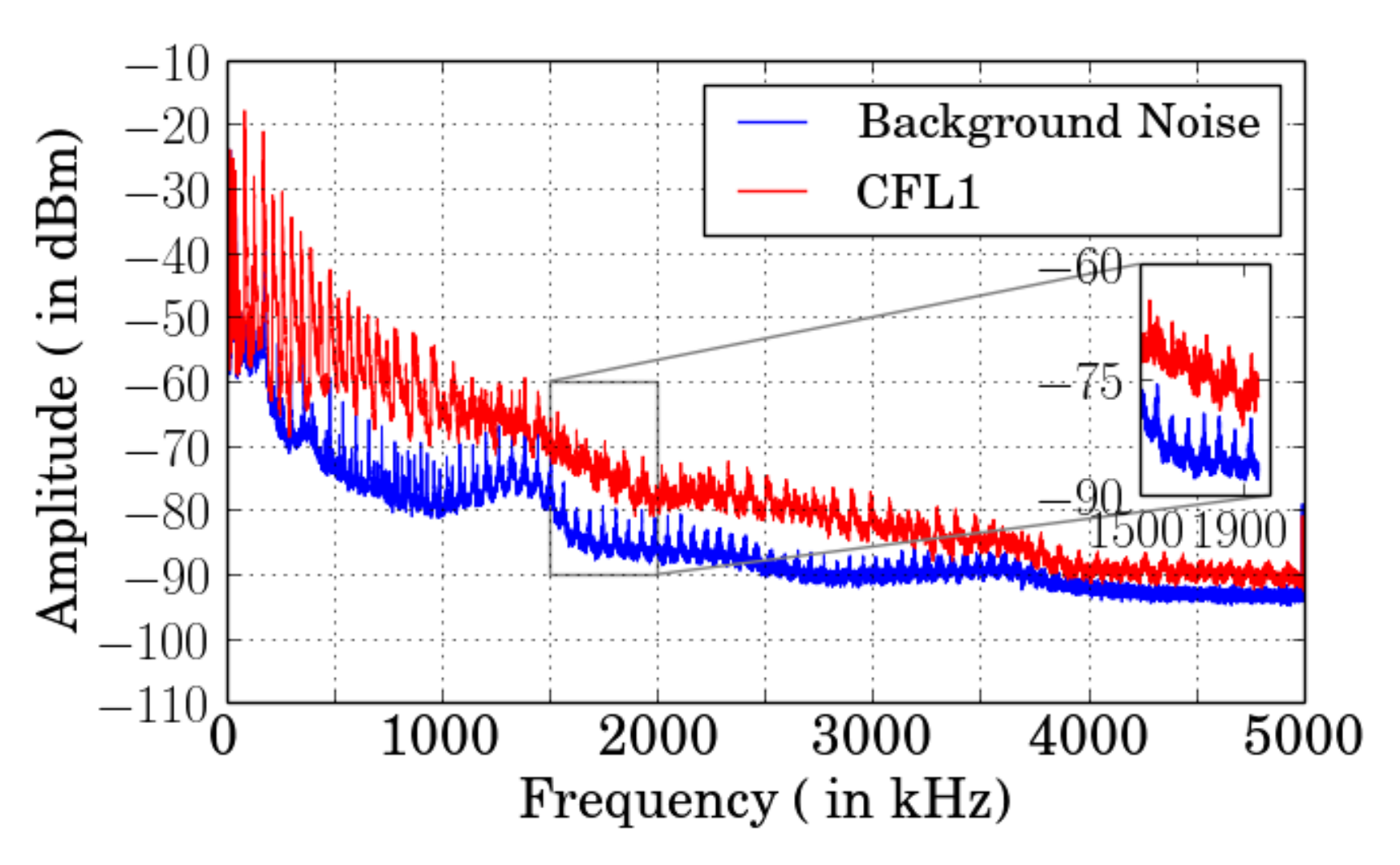}
    \caption{CFL1}
    \label{fig:obs1_1}
    \end{subfigure}%\hspace{70pt}%\hfill
    \begin{subfigure}{0.49\columnwidth}
    \includegraphics[width=\textwidth]{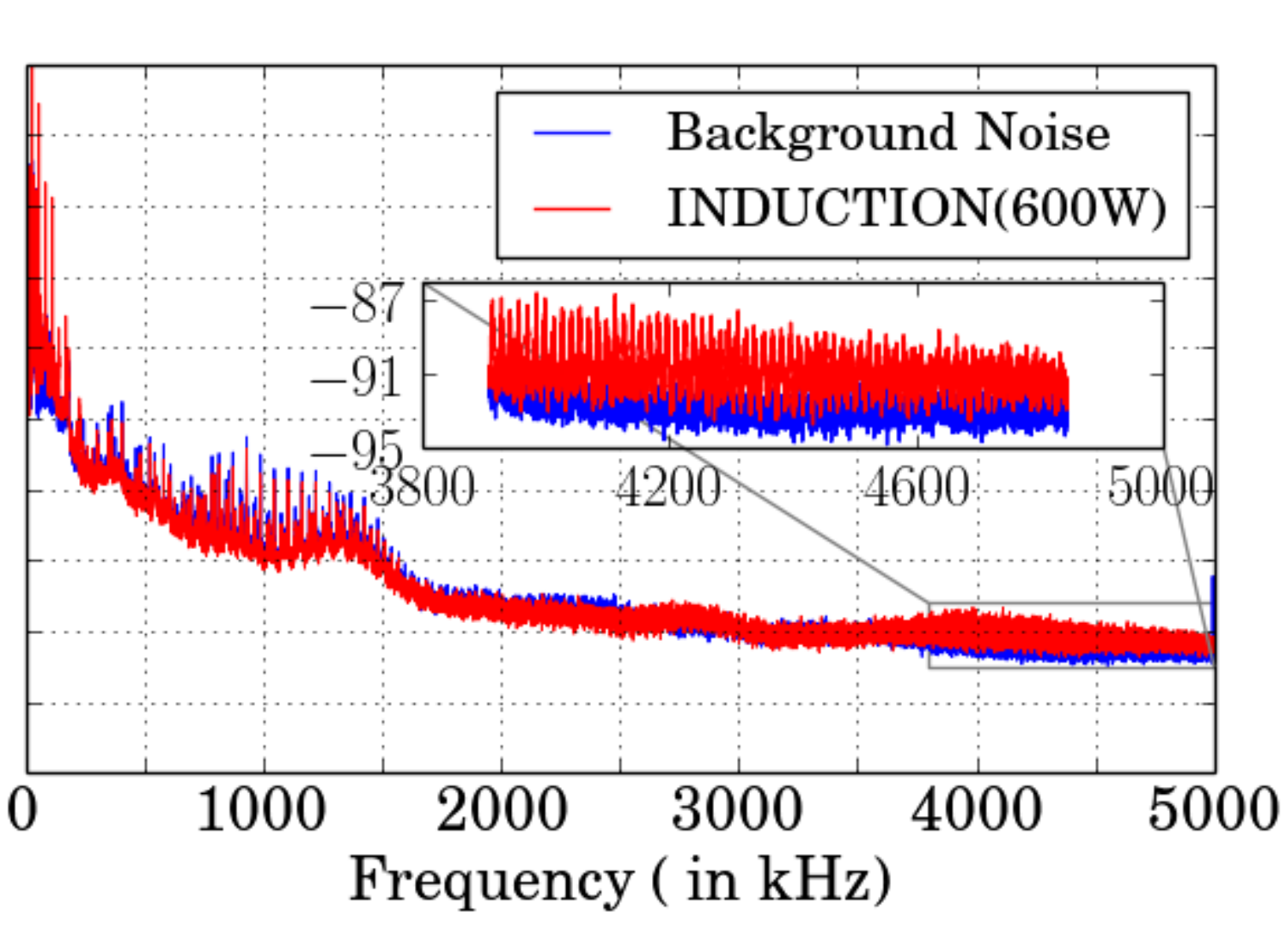}
    \caption{Induction Cooktop}
    \label{fig:obs1_2}
    \end{subfigure}%\hspace{51pt}%\hfill%
    \begin{subfigure}{0.5\columnwidth}
    \includegraphics[width=\textwidth]{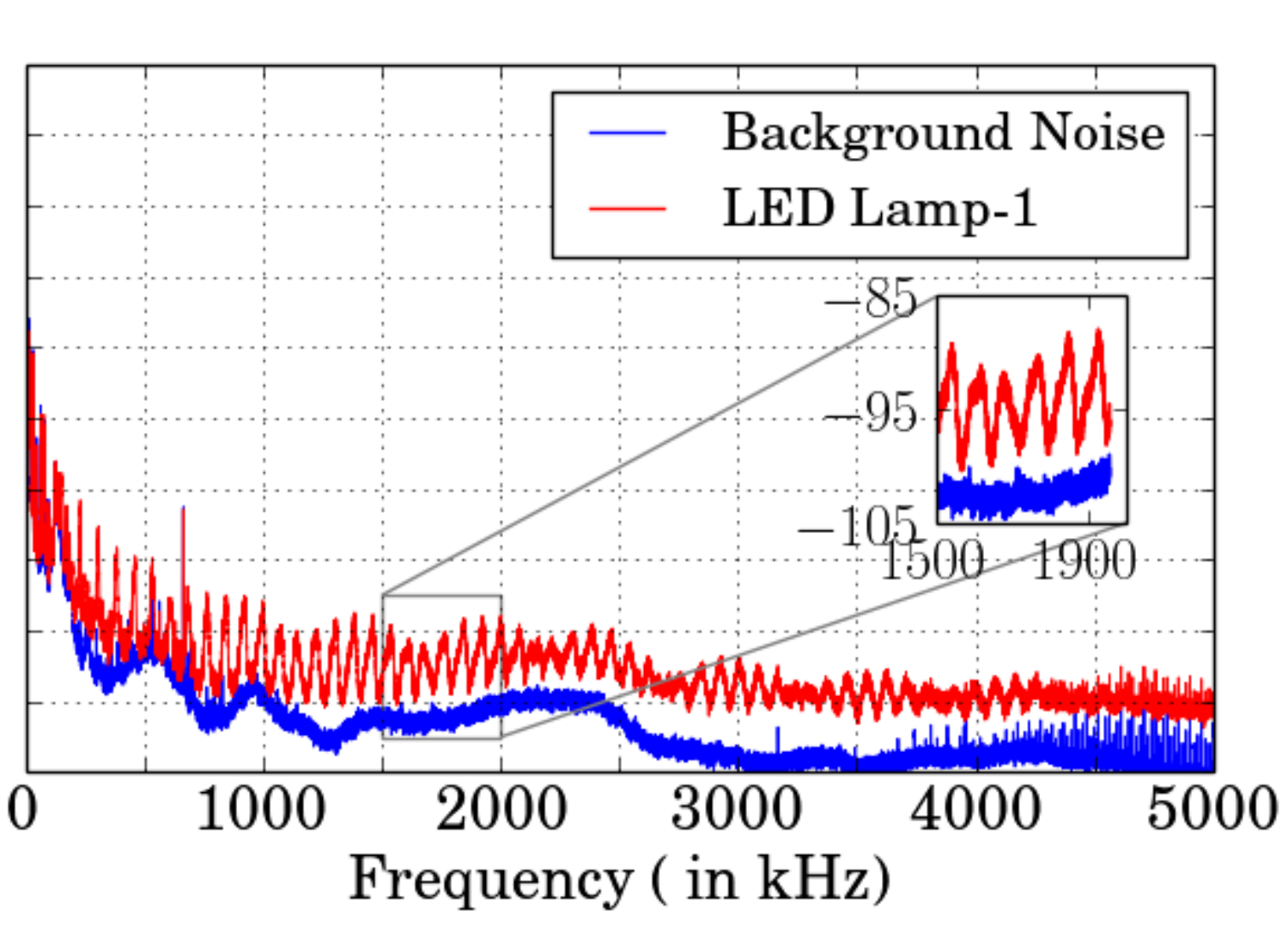}
    \caption{LED Lamp}
    \label{fig:obs1_3}
    \end{subfigure}%\hspace{51pt}%\hfill%
    \begin{subfigure}{0.5\columnwidth}
    \includegraphics[width=\textwidth]{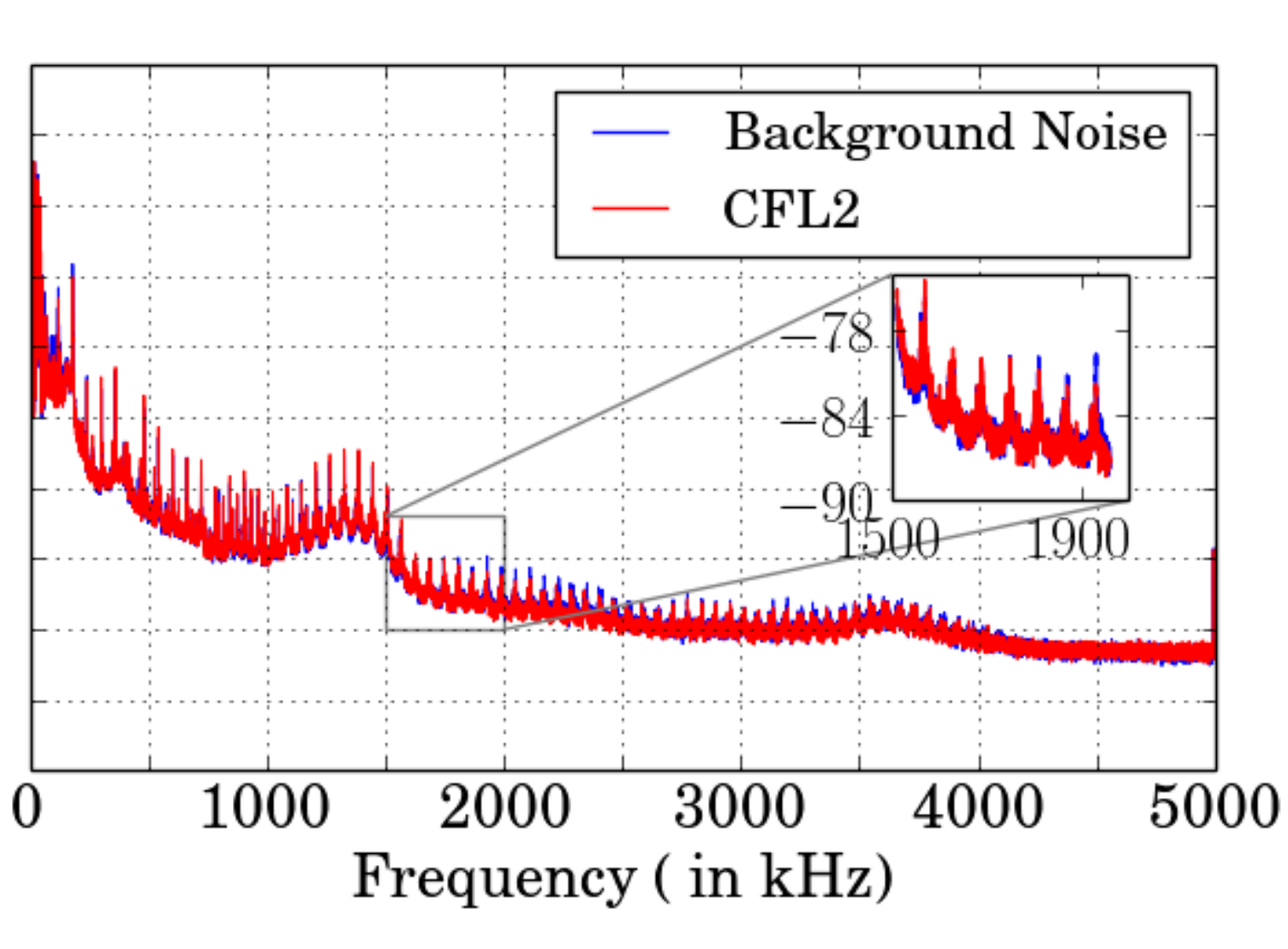}
    \caption{CFL2}
    \label{fig:obs1_4}
    \end{subfigure}
    \vspace{-10pt}
    \caption{Measured EMI from different AUTs connected independently to mains power supply along with background noise traces measured with AUTs disconnected from the setup.}
    \label{fig:obs1}
\end{figure*}

\subsection{Appliance Specific EMI Signatures}

\begin{figure*}[t!]
\vspace{-10pt}
\hspace{-7pt}
  \begin{subfigure}{0.6\columnwidth}
  \includegraphics[width=\textwidth]{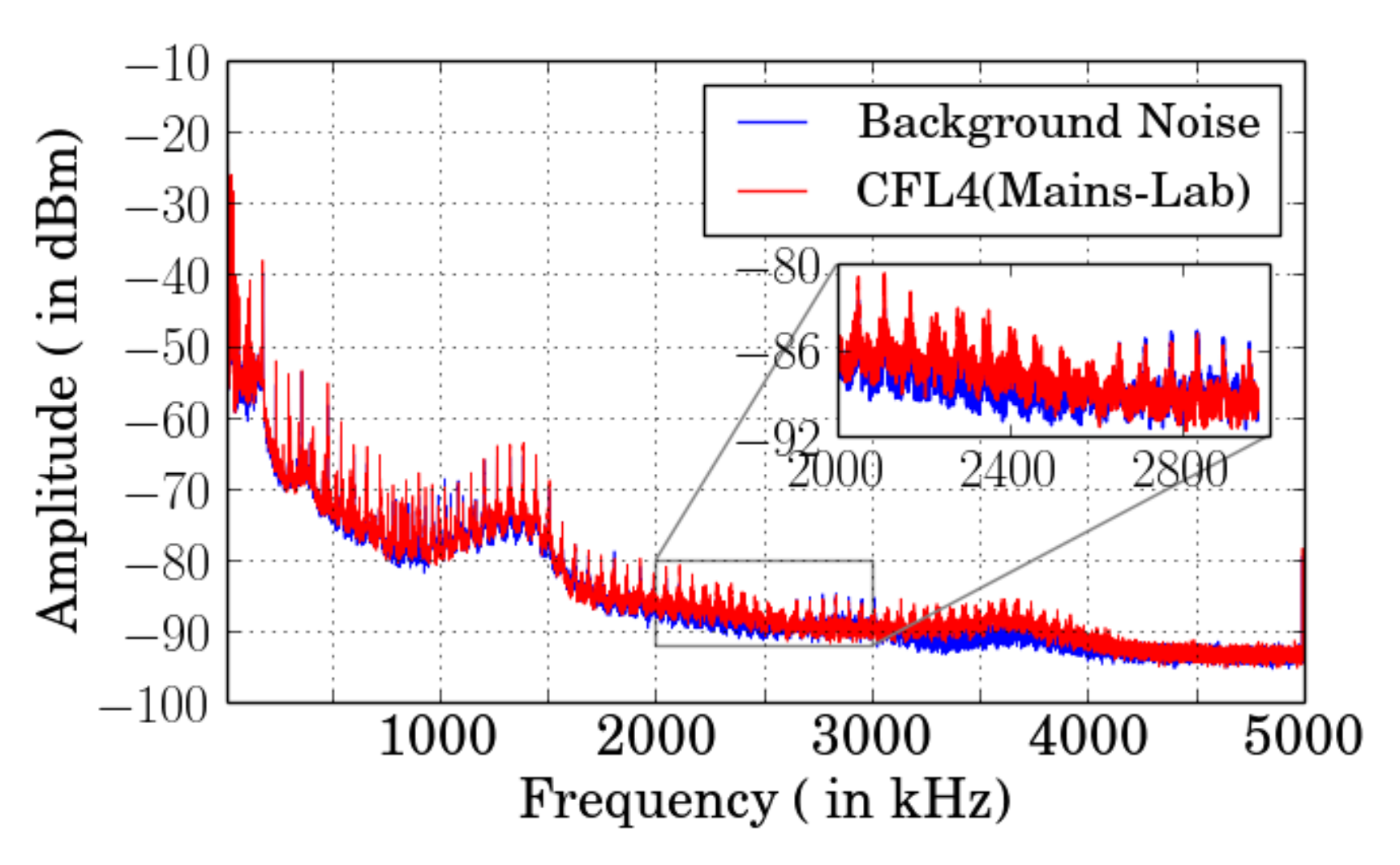}
  \caption{CFL4 (mains)}
  \label{fig:obs2_1}
  \end{subfigure}%\hspace{70pt}%\hfill
  \begin{subfigure}{0.5\columnwidth}
  \includegraphics[width=\textwidth]{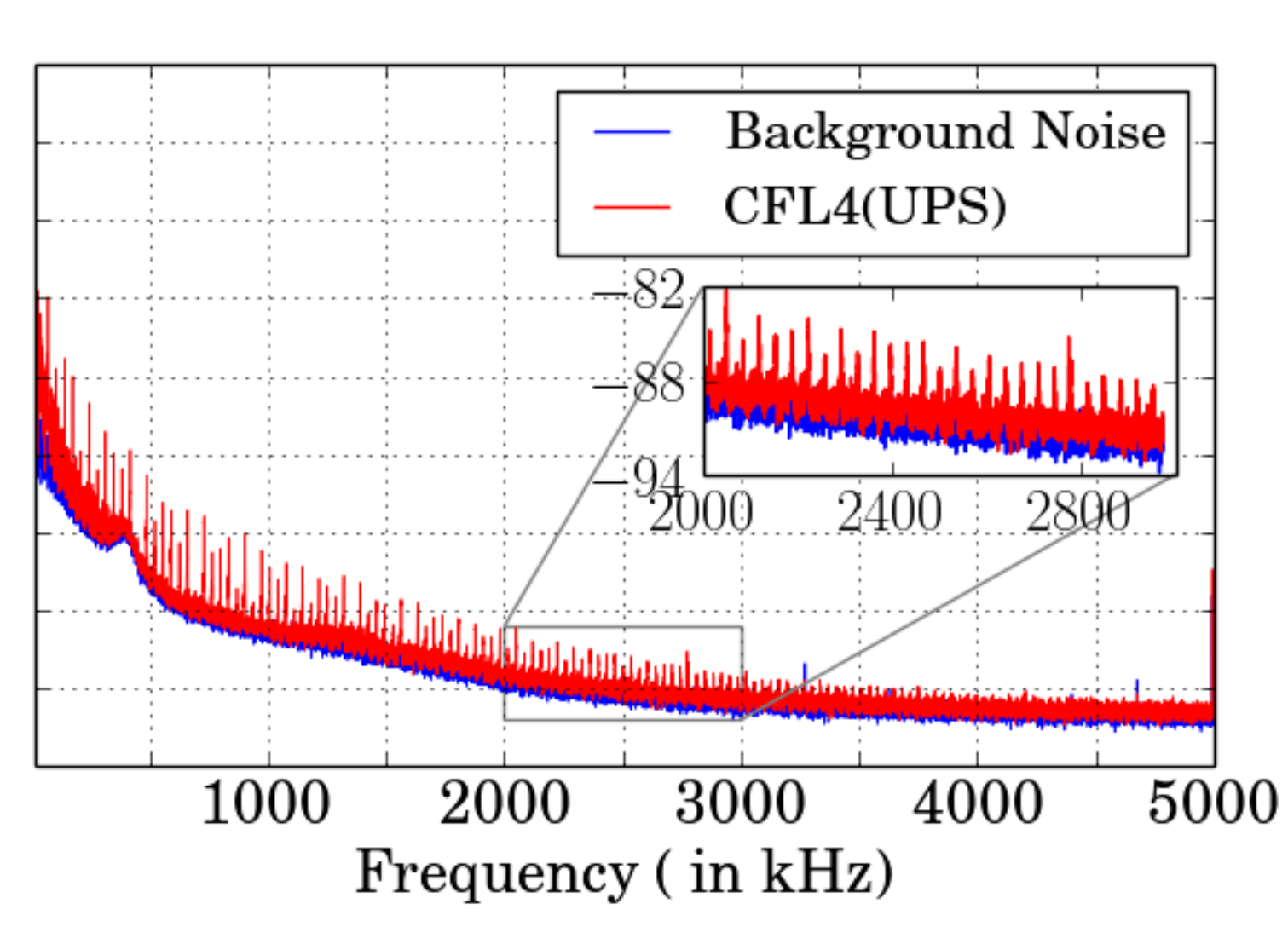}
  \caption{CFL4 (UPS)}
  \label{fig:obs2_2}
  \end{subfigure}%\hspace{51pt}
  \begin{subfigure}{0.5\columnwidth}
  \includegraphics[width=\textwidth]{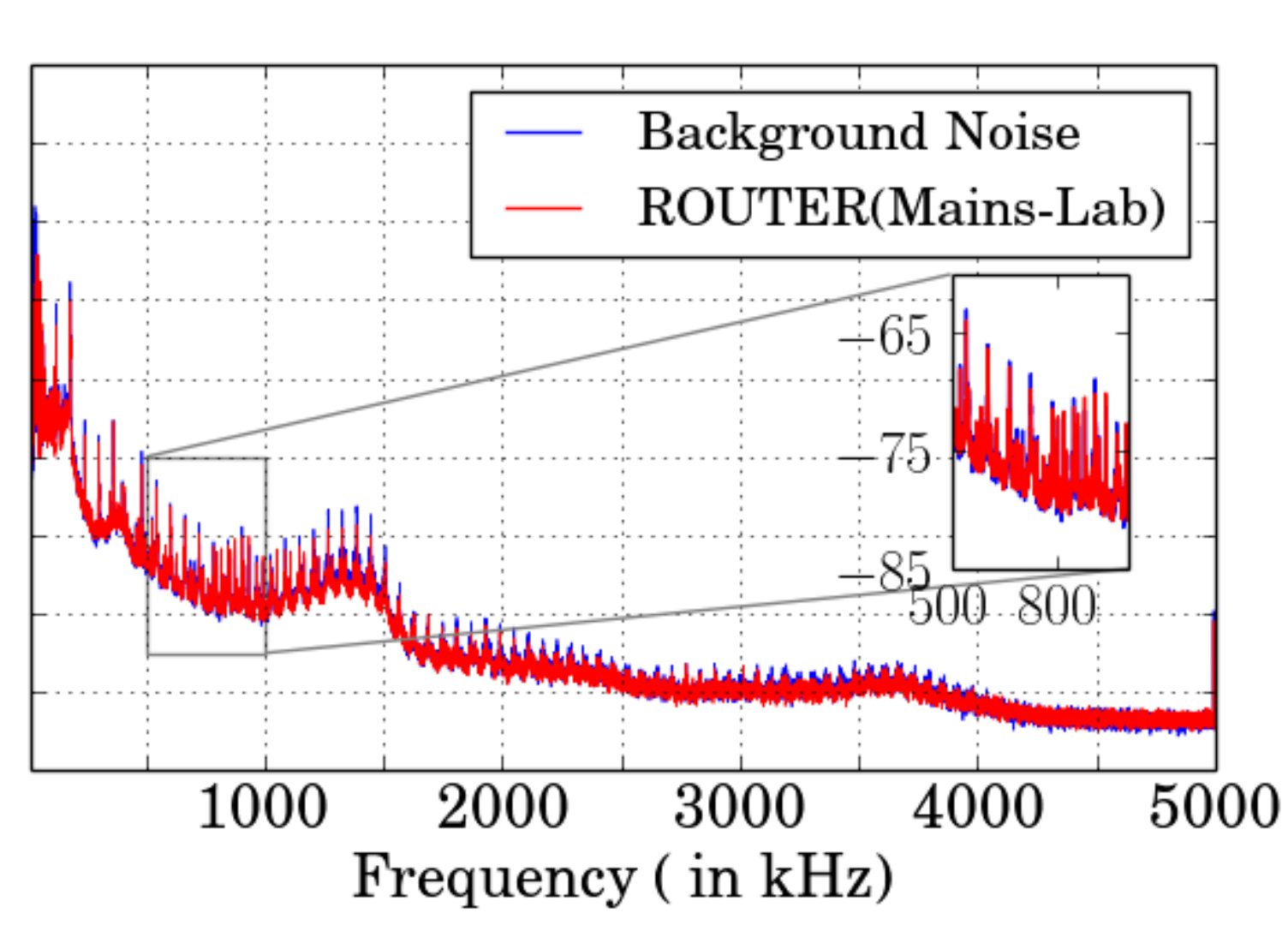}
  \caption{Router (mains)}
  \label{fig:obs2_3}
  \end{subfigure}%\hspace{51pt}
  \begin{subfigure}{0.5\columnwidth}
  \includegraphics[width=\textwidth]{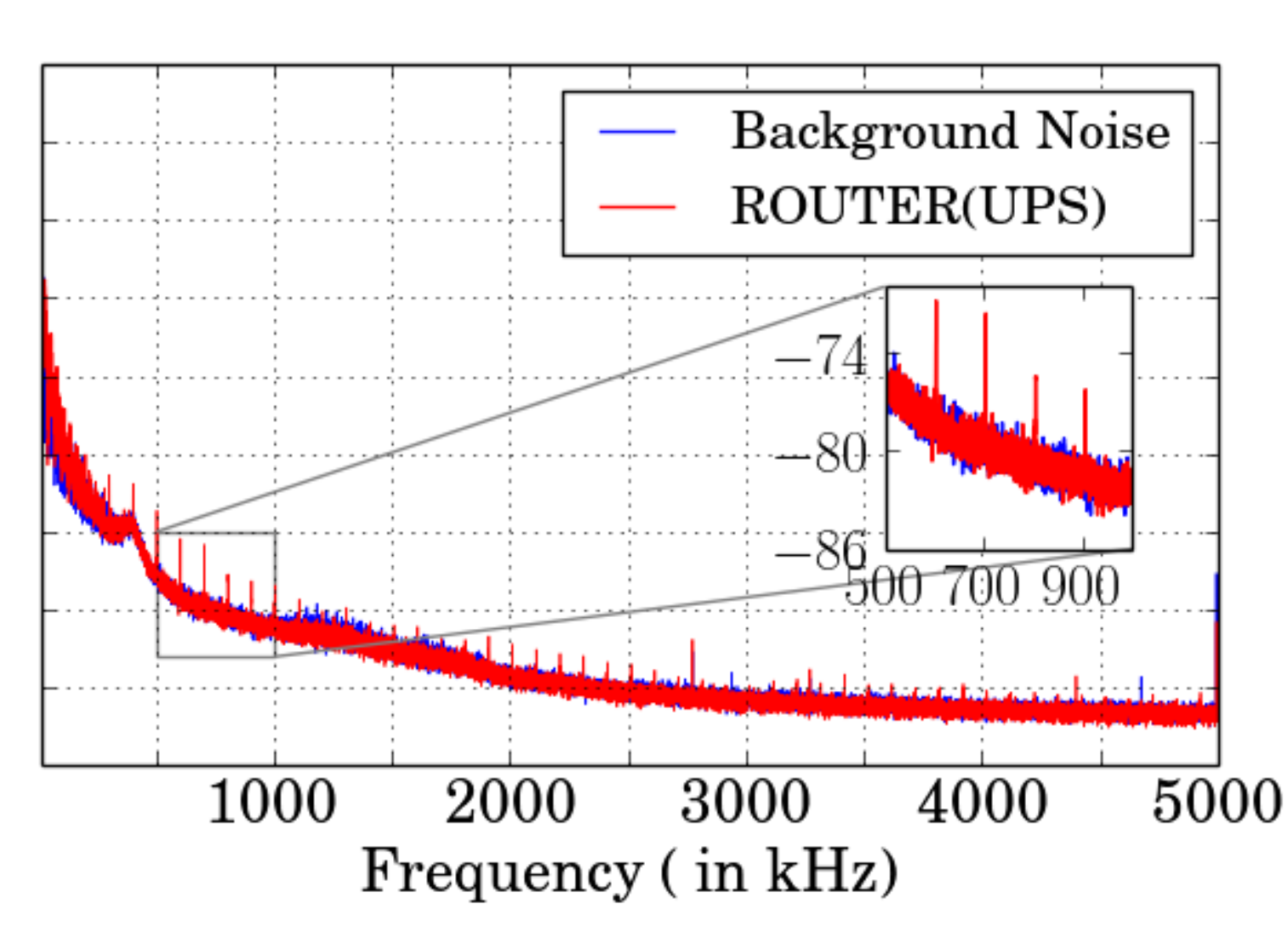}
  \caption{Router (UPS)}
  \label{fig:obs2_4}
  \end{subfigure}
  \vspace{-10pt}
  \caption{Background noise from mains interferes with EMI from (a)CFL4 and (c)router when they are connected to the mains power supply. UPS has much lower background interference as it is independently powered. Therefore, EMI from (b)CFL4 and (d)router can  be observed when they are connected to the UPS.}
  \label{fig:obs2}
\end{figure*}

In Figure-\ref{fig:obs1}, we display the EMI measurements taken from four AUTs, from test setup-1, along with background noise from the power line. Each appliance conducts EMI in a specific band of frequencies. CFL1 (Figure-\ref{fig:obs1_1}) conducts high EMI up to -25 dBm from 80 kHz to 3 MHz; the induction cooktop (Figure-\ref{fig:obs1_2}) conducts EMI from 3.5 MHz to 5 MHz; the LED lamp (Figure-\ref{fig:obs1_3}), conducts EMI from 1 MHz to 4 MHz; CFL2 (Figure-\ref{fig:obs1_4}) does not show any noticeable EMI. The EMI data from CFL1, CFL2 and induction cooktop were collected in the lab while the LED lamp data was collected in the home. The background noise traces clearly indicate that the harmonics of the 50Hz power signal are suppressed by the high pass filter. Also, the background noise level in the lab data was consistently higher than the background noise level at home. We believe this is because of the complex electrical infrastructure in the building where our laboratory is located. Besides CFL2, other appliances such as CFL3 and laptop chargers 1\&2 did not conduct any observable EMI. These results provide two useful insights - (1) All SMPS based appliances may not conduct significant EMI; (2) EMI signatures of the same class of appliances such as CFLs (Figures-\ref{fig:obs1_1}, \ref{fig:obs1_4} and \ref{fig:obs2_1}) may not be consistent across different manufacturers. 
% Observation-2 : Effect of BGN

\subsection{Effect of Background Noise}

Two appliances - a CFL and a router, are used to examine the effect of background noise on the measurable EMI from an AUT. Figure~\ref{fig:obs2} illustrates observed EMI for both the AUTs when connected to mains power supply (setup-1) and UPS (setup-4). The EMI from both the appliances are hardly noticeable when they are connected to the mains power supply (see Figure-\ref{fig:obs2_1} for CFL4 and \ref{fig:obs2_3} for router). This is because of the high interference feature arising from the background appliances on the same power line, resulting in poor signal to interference ratio. When connected to an independently powered UPS, that is isolated from the electrical infrastructure of the building, both the appliances show significant EMI (see Figure-\ref{fig:obs2_2} and \ref{fig:obs2_4} for CFL4 and router respectively). The thermal noise floor is uniform in both the cases and is limited by the bandwidth of the measurement device. These results indicate that in certain sensing configurations, with high background interference, disaggregation of individual appliances with low EMI may be challenging.

\subsection{Effect of Power Line Impedance}

\begin{figure}[t!]
%\centering
\vspace{-10pt}
    \hspace{+20pt}
%    \begin{subfigure}{0.85\columnwidth}
    \includegraphics[height=4.5cm]{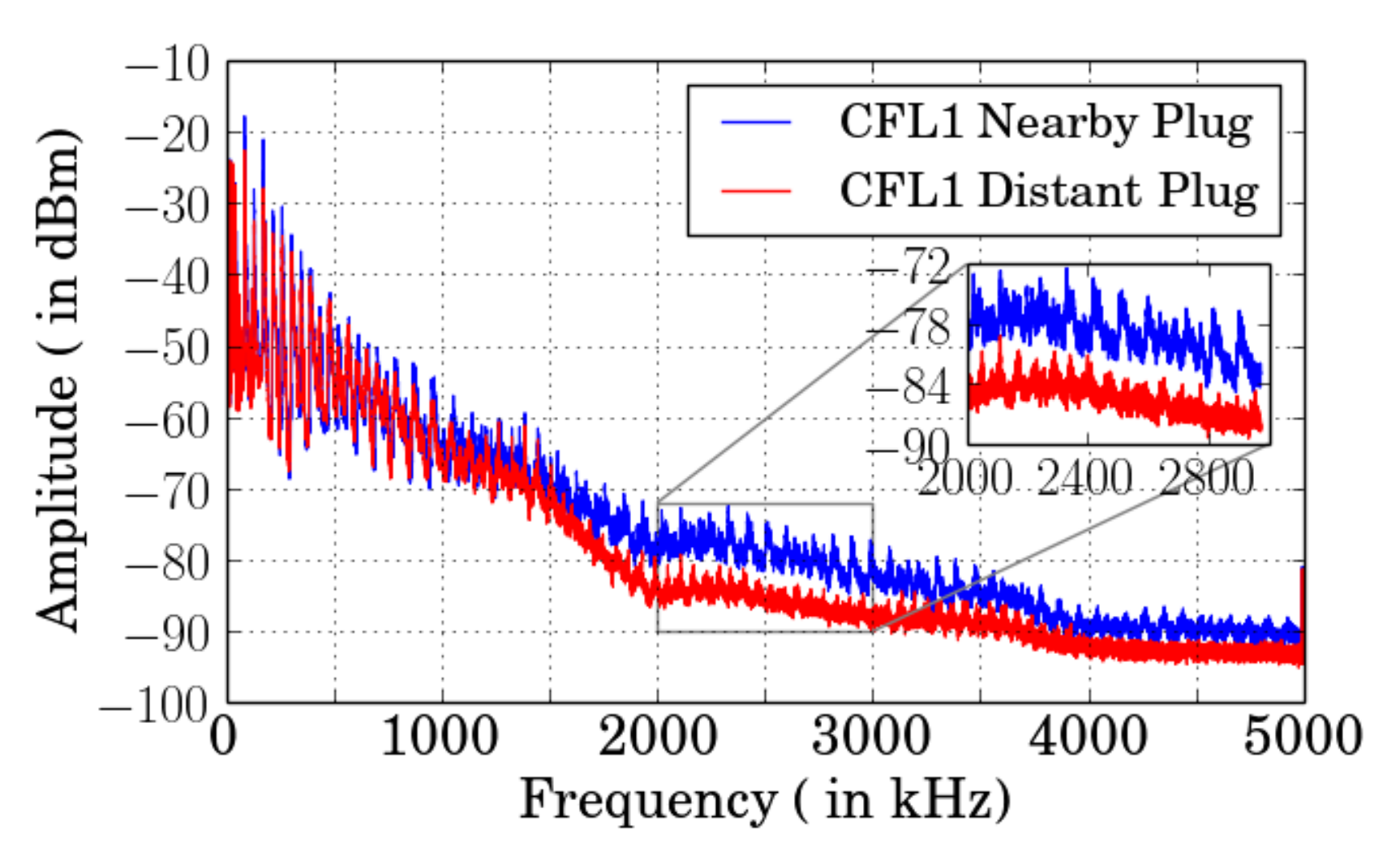}
%    \vspace{-20pt}
%    \label{fig:obs4_1}
 %   \end{subfigure}%\hspace{70pt}
    % \\
    % \begin{subfigure}{0.85\columnwidth}
    % \epsfig{file=ROUTER_SPP_DPP.eps,width=\textwidth}
    % \vspace{-20pt}
    % \caption{}
    % \label{fig:obs4_2}
    % \end{subfigure}%\hspace{51pt}%\hfill%
\vspace{-10pt}
\caption{Effect of line impedance on EMI from CFL1 on a nearby and a distant plug point}
\label{fig:obs4}
\end{figure}

Next we study the impact of power line impedance on the EMI conducted by AUTs. We measure the EMI trace of an AUT, connected at a near plug point (NPP) from the sensing plug point (SPP) where the measurement device is connected (setup-1). Then the measurement is repeated with the AUT connected at a distant plug point (DPP), 10 meters from the SPP (setup-2). Figure-\ref{fig:obs4} shows that EMI from CFL1 was attenuated at DPP in comparison to EMI observed at NPP especially at high frequencies above 2 MHz. This can be attributed to the transmission line effects of the long extension cord. 
An in-depth analysis of higher order effects of the transmission line is required to better understand the impact of line impedance on EMI measurements.
% Observation-3
\begin{figure*}[t!]
    \begin{subfigure}{0.34\textwidth}
    \includegraphics[width=\textwidth]{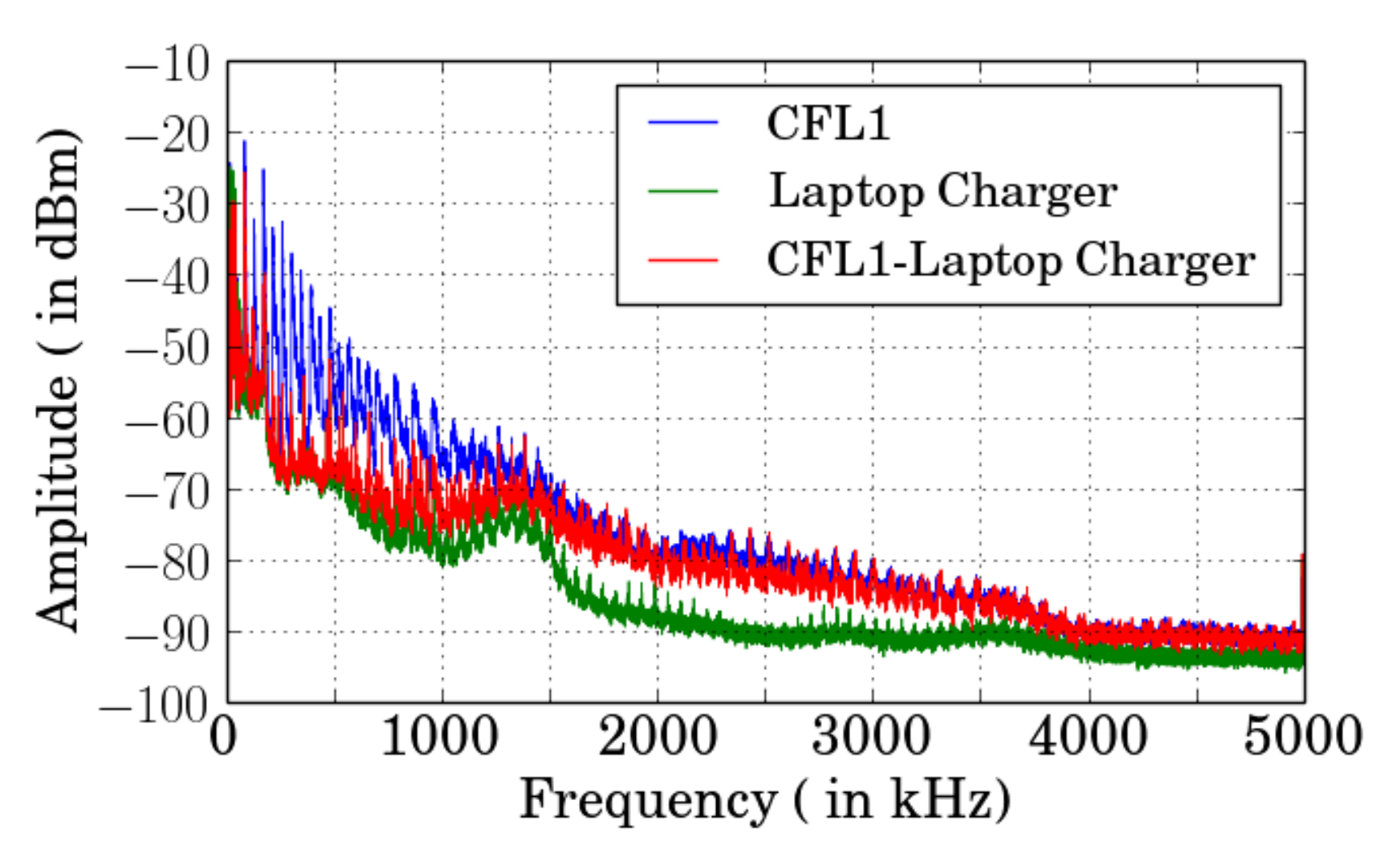}
    \vspace{-10pt}
    \caption{CFL and Laptop Adapter}
    \label{fig:obs3_1}
    \end{subfigure}\hspace{-6pt}
    \begin{subfigure}{0.34\textwidth}
    \includegraphics[width=\textwidth]{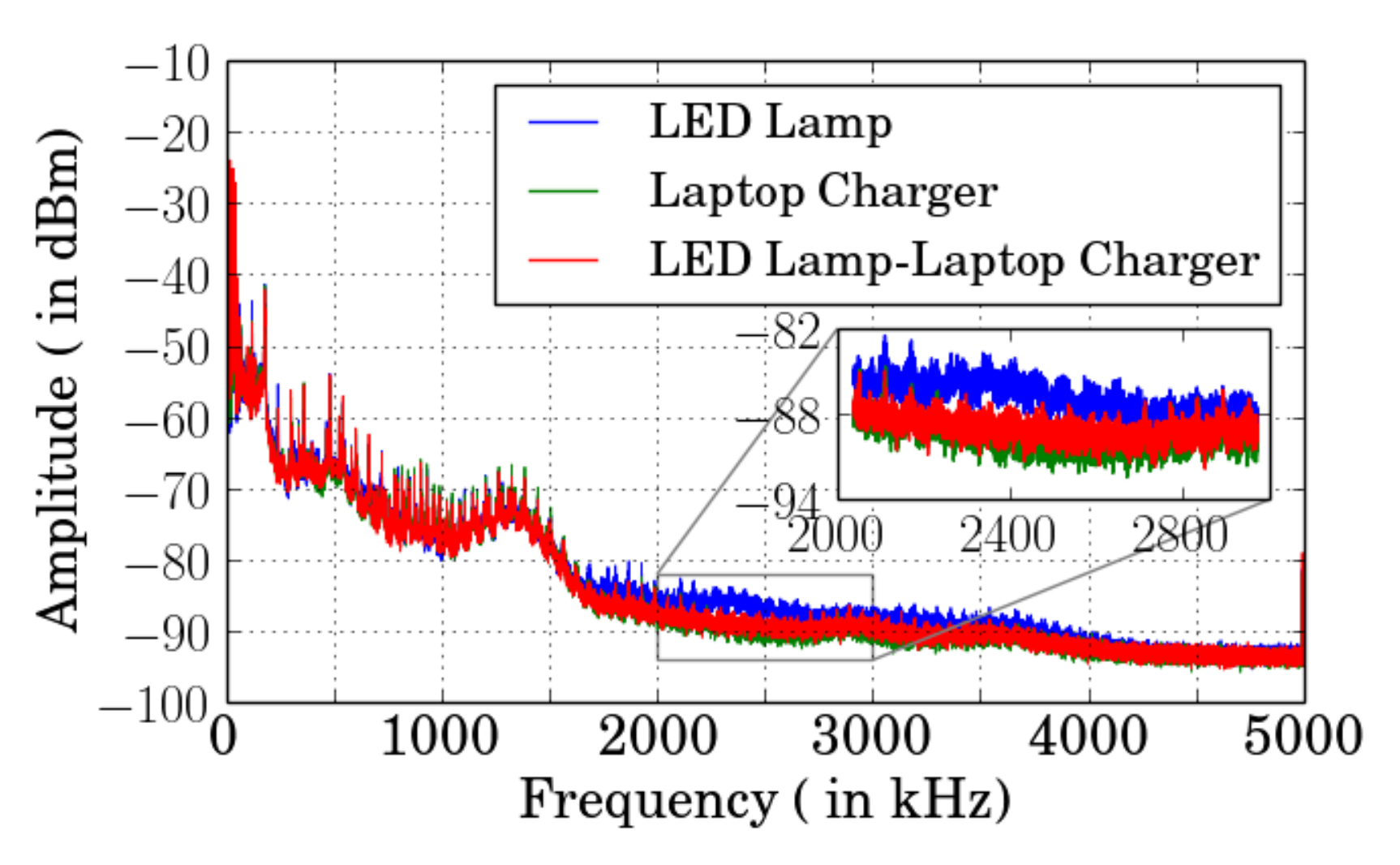}
    \vspace{-10pt}
    \caption{LED and Laptop Adapter}
    \label{fig:obs3_2}
    \end{subfigure}\hspace{-6pt}
    \begin{subfigure}{0.34\textwidth}
    \includegraphics[width=\textwidth]{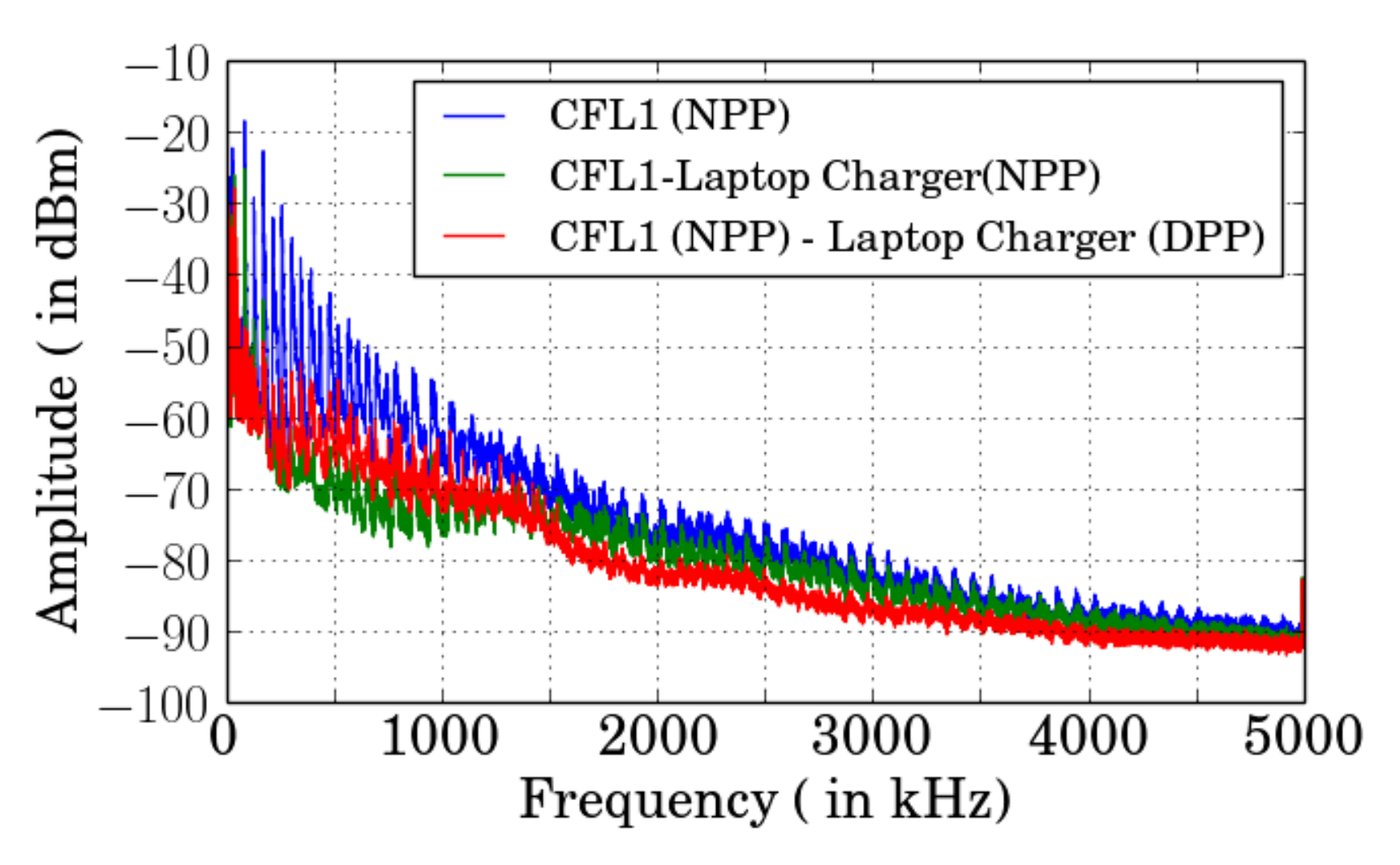}
    \vspace{-10pt}
    \caption{CFL (NPP) and Laptop Adapter (DPP)}
    \label{fig:obs3_3}
    \end{subfigure}%\hspace{51pt}
\vspace{-10pt}
\caption{EMI as observed with a combination of CFL1 and LED lamp with laptop adapter on the same power line.}
\label{fig:obs3}
\end{figure*}

% Observation-4 Effect of line impedance

% Observation-5
\begin{figure}[t!]
\vspace{-10pt}
\hspace{+20pt}
%\begin{subfigure}{0.5\columnwidth}
\includegraphics[height=4.5cm]{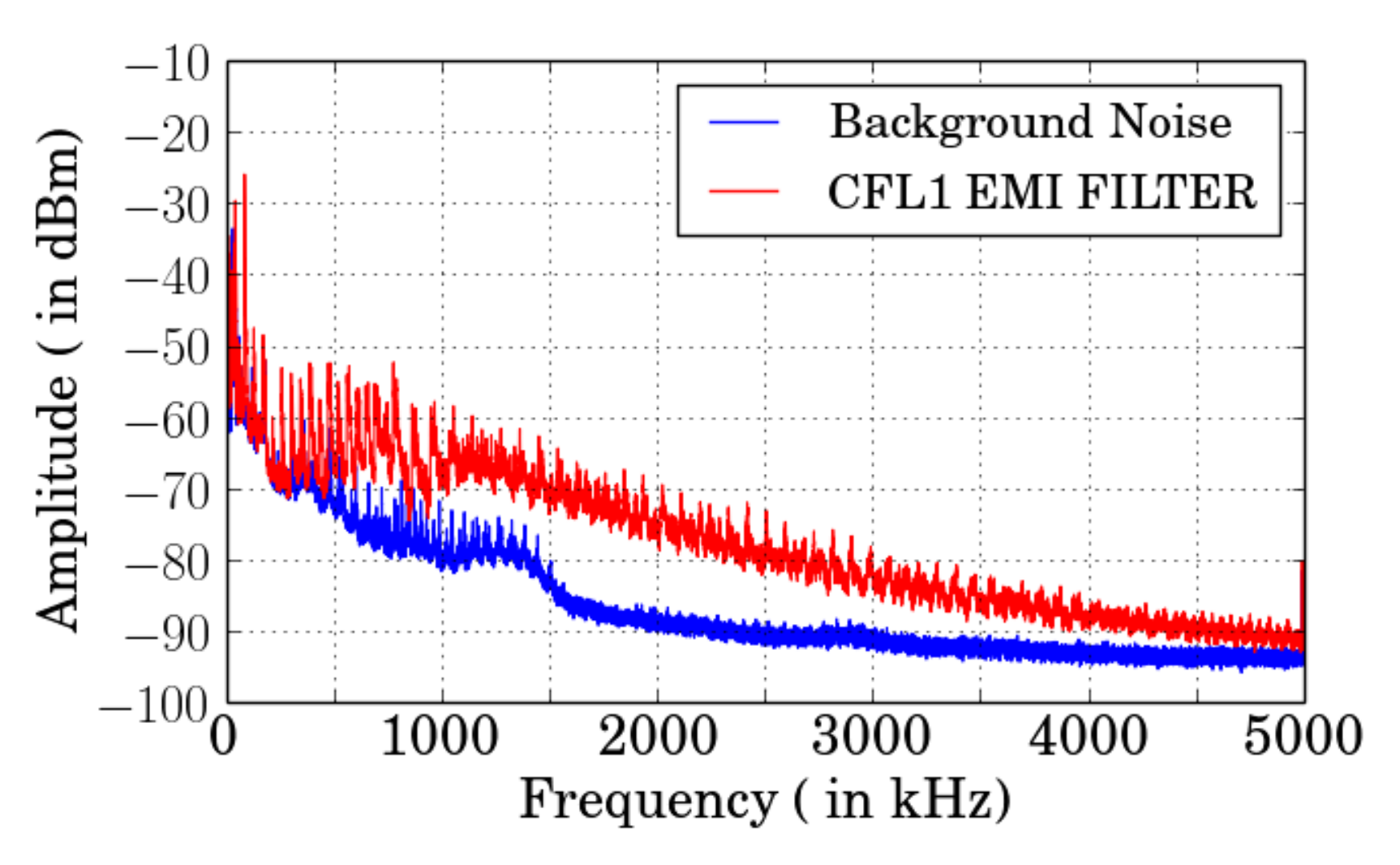}
%\end{subfigure}
\vspace{-10pt}
\caption{Impact of inbuilt EMI filter of the extension cord on the EMI observed from CFL1 connected to the mains power supply}
\vspace{-10pt}
\label{fig:obs5}
\end{figure}

Some extension cords have inbuilt EMI filters, to protect the appliances from noise in the power supply. In order to study their impact, we measured the EMI from CFL1 when an external EMI filter is connected between the off-the-shelf extension cord and mains power supply (setup-3). This result is shown in Figure-\ref{fig:obs5} and compared with the EMI, shown earlier in Figure-\ref{fig:obs1_1}, from setup-1 which did not have an EMI filter on the extension cord. We observe two effects of the power line filter. First, the strength of background interference features from the power supply is lowered. Second, the EMI from the AUT is also lowered. We postulate that this phenomenon occurs because EMI currents from the AUT flow into the low impedance path of the filter and hence cannot be measured at the sensing point. Thus the power line EMI filters impact the measurability of EMI, of an appliance connected to the power line.

\subsection{Effect of Multiple Appliances on a Common Power Line}

Finally, we consider the case, when multiple AUTs are connected on the same power line. This is a common setting in a home where different appliances may operate simultaneously, during some time intervals. Figure-\ref{fig:obs3_1} shows the EMI conducted by CFL1 and laptop charger when powered independently and when connected together. As mentioned earlier - when connected independently, the EMI from the CFL1 is clearly observed while the EMI from the laptop charger is weak due to its inbuilt EMI filter. Interestingly, when the two appliances are connected together, the EMI of the CFL1 is suppressed from 40 kHz to 1.5 MHz. 

We postulate that this phenomenon is the result of the low impedance path offered, to the EMI currents from CFL1, by the inbuilt EMI filter in the laptop charger. Similar behavior is observed when an LED lamp and a laptop charger are connected together (Figure-\ref{fig:obs3_2}). We further investigate this phenomenon, by measuring the EMI (as shown in Figure-\ref{fig:obs3_3}) when the CFL1 is connected to NPP with respect to SPP, while the laptop charger is connected at DPP. In this case, the EMI currents from CFL1 do not flow through the filter in the laptop charger and instead flow through the sensing point, where they can be measured. This is due to the line impedance introduced between the two AUTs. The damping effect of the inbuilt EMI filters of an appliance on the EMI signatures of neighbouring appliances on the power line is significant only when the appliances are in close proximity. 

\section{Simulation Models for EMI}
    \label{sec:EMImodel}

During EMI measurements, we observed certain factors that affected the conducted EMI from an appliance, such as line impedance and appliance coupling behavior (specifically for appliances with low pass EMI filters). We now propose a generic model for EMI that can be customized to mimic actual setup to gain a better understanding of these observations. There are three main objectives towards developing a detailed simulation model: (1) Actual measurements of EMI from an appliance are complicated due to the presence of several appliances in the background. Simulation models allow us to independently analyze the EMI from an individual appliance as well as combinations of multiple appliances; (2) Simulations allow us to analyze the impact of transmission line parameters of the mains power line which are difficult to measure in real world settings; (3) Simulation models can be, potentially, useful for generating large volumes of training databases for supervised learning techniques, for appliance level disaggregation.

We, first, give an overview of SMPS operation, which forms the basis of our simulation models.  Typically, in residential settings, alternating current (AC) to direct current (DC) step-down SMPS topologies interface between an appliance and the power line. The primary section of these SMPS consist of a transformer where the 230VAC signal is stepped down to a low voltage AC signal and subsequently converted to an unregulated DC signal with a rectifier. The rectified DC is fed to a DC to DC converter, with a high-speed switching circuit, to generate a regulated DC output that powers the appliance load or its control sections. The switching frequencies of these circuits rely upon the output power requirements of the appliance and hence may be unique to an appliance \cite{boost2007switch}. Due to the highly nonlinear nature of these switching circuits, the dominant component of EMI from an appliance consists of the switching frequency and its harmonics. The EMI currents are coupled to the mains power line through galvanic, inductive or capacitive coupling modes, within the circuit \cite{paul2006introduction}. In this paper, we focus on simulating the EMI currents generated in the DC to DC converter section of the SMPS. We assume that the coupling modes of the EMI to the power line are fairly uniform across multiple appliances and do not significantly alter the nature of the EMI spectrum, at frequencies up to 5 MHz. 
% Equivalent Model for EMI
\begin{figure}[t!]
  \vspace{-10pt}
  \hspace{20pt}
  \includegraphics[height=4cm]{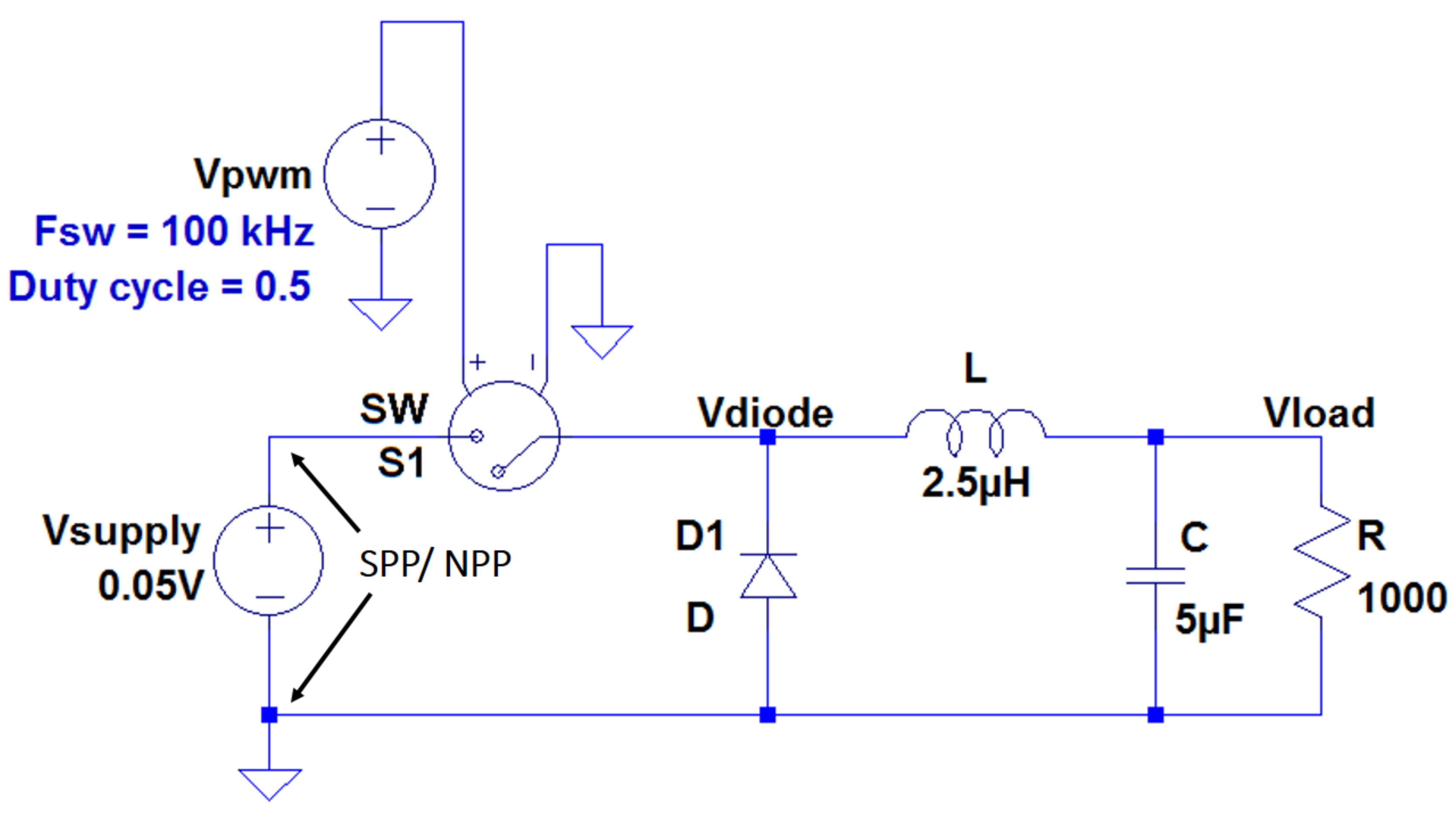}
  \vspace{-10pt}
  \caption{Spice model of a simplified buck convertor used as an equivalent appliance model for EMI generation}
  \vspace{-10pt}
  \label{fig:Model}
\end{figure}

To simulate the behavior of conducted EMI, an equivalent model for an appliance's power supply is presented here, utilizing a simplified version of a buck converter shown in Figure-\ref{fig:Model}. A detailed description of the behavior of a buck converter is presented in \cite{boost2007switch}.   A buck converter is a standard DC to DC step down converter that utilizes a MOSFET or a transistor as a switch, S1, to generate a pulsating DC (Vdiode) from input unregulated DC (Vsupply). The switching frequency (Fsw) and the duty cycle of this pulsating DC are varied by a pulse width modulated (PWM) control signal driving S1. Vdiode is averaged using an inductor-capacitor (LC) combination to provide a constant DC (Vload) at the load. The diode, D1, provides a path for the current through the inductor to discharge when S1 is ``open". The ratio of Vload to Vsupply is governed by the duty cycle of the PWM control signal. A standard buck converter may also consist of a feedback loop to maintain a constant output voltage, in the case of a fluctuating input voltage. We omit this feedback loop in our simulations and, instead, assume a constant DC supply voltage. To simplify this model, we also assume that the load impedance of the appliance is purely resistive (R). The currents (Isupply) drawn from the DC supply of the DC to DC converter are coupled to the mains giving rise to EMI. Therefore, the frequency response of Isupply provides an EMI signature of the appliance. 

We model the EMI from a specific SMPS based appliance on the basis of EMI measurements of the appliance when powered by a UPS. The input parameters governing the simulated EMI spectrum are Vsupply, duty cycle and Fsw of PWM control signal, and load impedance (Z/R). Fsw is chosen based on the fundamental frequency component and harmonics observed in the measured EMI spectrum. The baseline of the simulated EMI spectrum is adjusted to match the noise floor in the measured EMI spectrum by modifying the value of Vsupply. Vload and R, are adjusted to control the amplitude of Isupply, which in turn, controls the magnitude of EMI peaks in the frequency spectrum. The component values for L \& C are computed by buck converter design equations \cite{boost2007switch}. The width of the EMI peaks, observed in the frequency spectrum, are adjusted by introducing suitable series resistances to L and C. The main advantage of the simulation method that we have proposed is that, our model is created on the basis of accurate EMI measurements of an appliance and not on any prior knowledge of the internal circuitry of the appliance which is usually difficult to obtain.
% Time domain behavior of EMI Model
\begin{figure}[t!]
  \vspace{-10pt}
  \includegraphics[height=4.2cm]{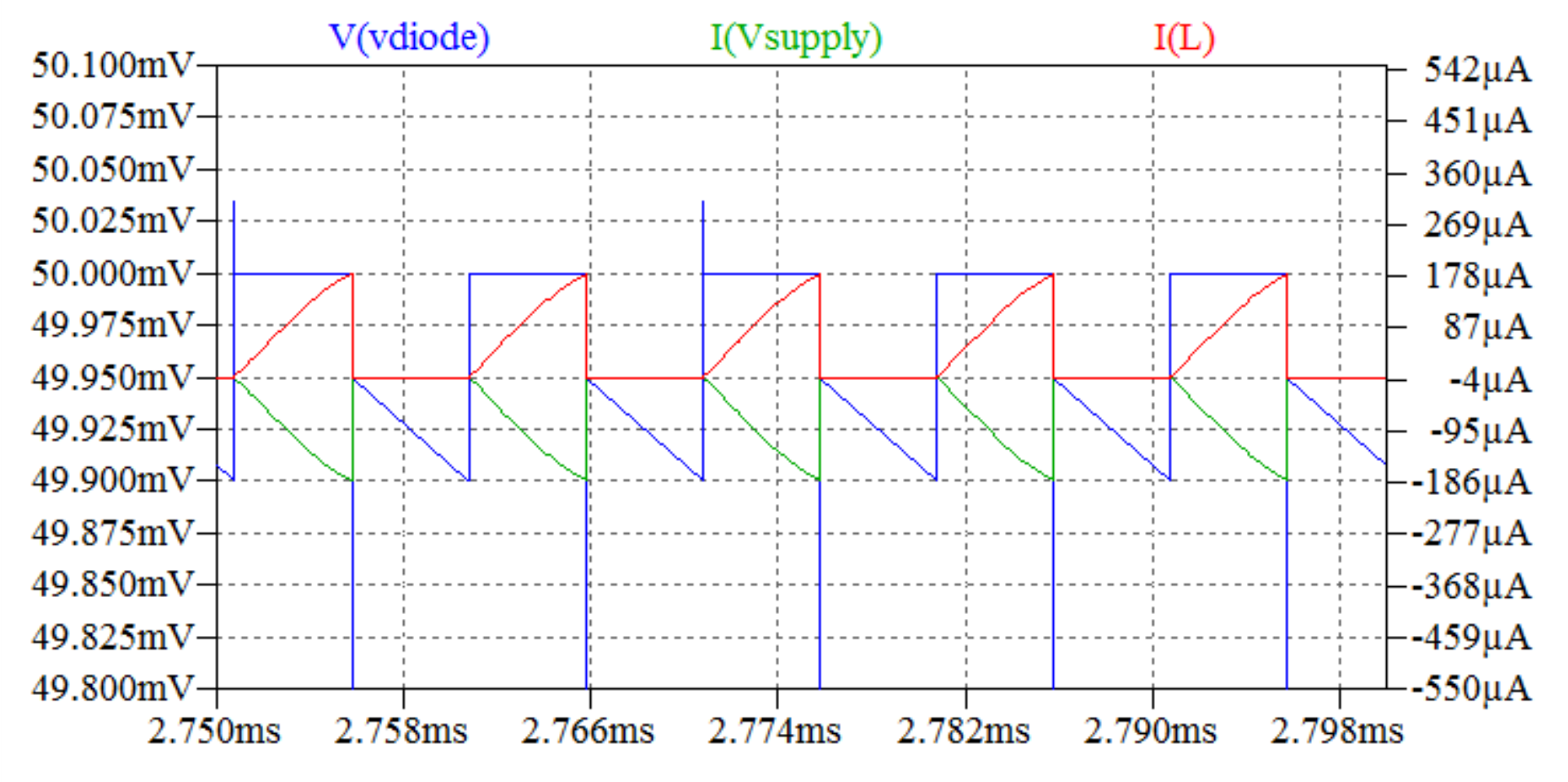}
  \vspace{-10pt}
  \caption{Simulated steady-state time domain behavior of a simplified buck convertor used as an equivalent appliance model. I(Vsupply) is the EMI current that couples to the mains power line.}
  \vspace{-10pt}
  \label{fig:timedomain}
\end{figure}

% Freq. Domain results from simulation and measurements
\begin{figure}[t!]
  \hspace{20pt}
  % \begin{subfigure}{0.5\columnwidth}
  \includegraphics[height=4.2cm]{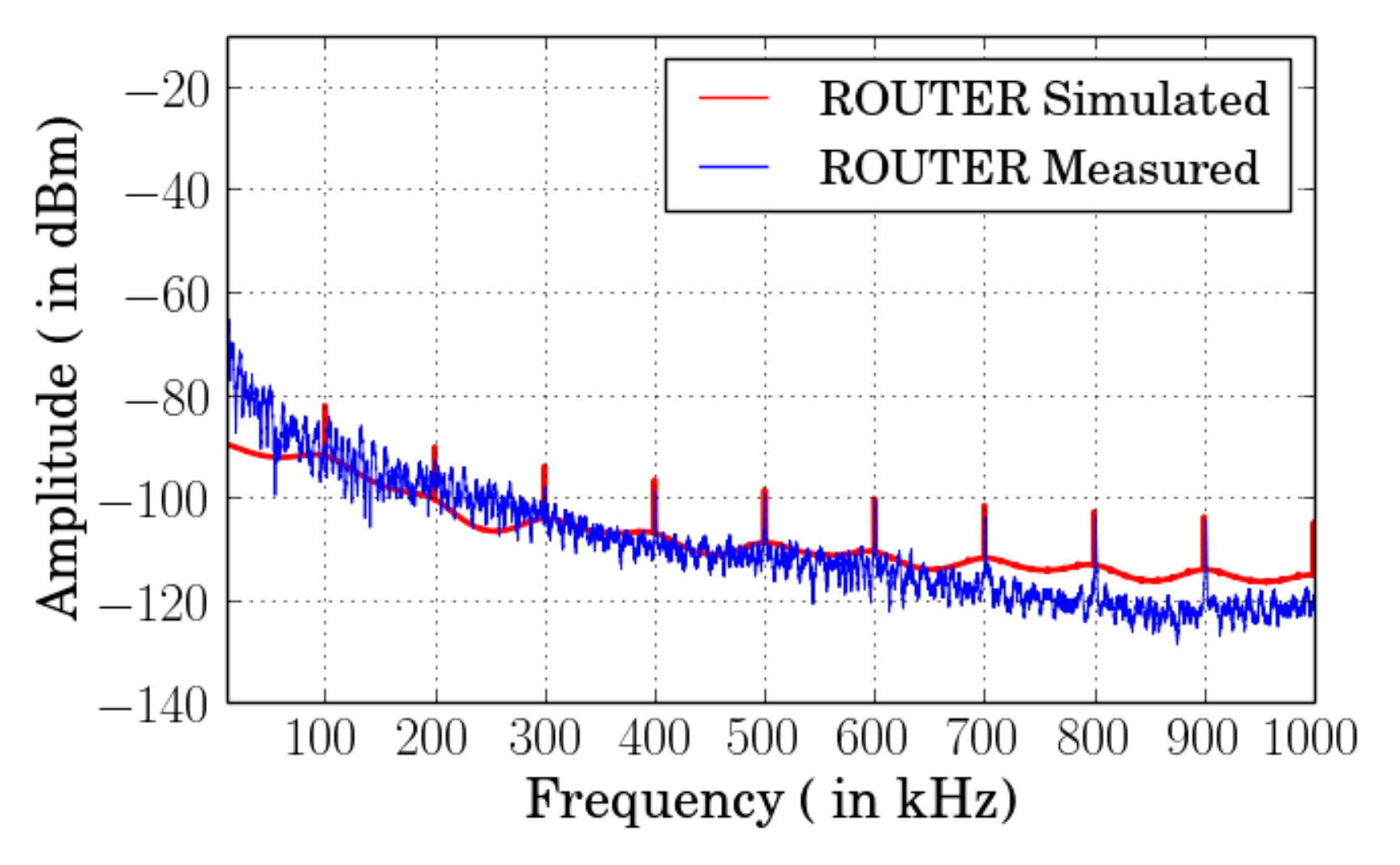}
  % \end{subfigure}
  \vspace{-10pt}
  \caption{Frequency spectrum of simulated EMI from I(Vsupply) along with EMI observed from router on test setup-3}
  \vspace{-10pt}
  \label{fig:freqdomain}
\end{figure}

\subsection{Appliance Specific EMI}

We present a model to simulate the EMI spectrum of a router based on measurements, presented earlier in Figure-\ref{fig:obs2_4}.  We computed the following design parameters: Vsupply=0.05V, duty cycle=0.5, Fsw=100 kHz, L=2.5uH, C=5uF and R=1000$\Omega$.

Figure-\ref{fig:timedomain} illustrates the steady-state time-domain results from the simulations. V(vdiode), the pulsating DC at the diode,  has a time period of 1/Fsw and duty cycle of 0.5. I(L), the current across the inductor, continually charges and discharges during the ON time and OFF time of the pulse. I(Vsupply) is the current drawn from the DC power supply and is highly non-linear as indicated by the inverted triangular shaped pulses. This current is the primary component that couples to the mains power supply giving rise to conducted EMI. The frequency domain response of I(Vsupply) is presented in Figure-\ref{fig:freqdomain}. Qualitatively, the simulated and measured spectrums show a good match. For instance, the EMI peaks, the baseline and the width of the frequency peaks are similar in both the spectrums. Although the frequency spectrum shows spectral overlap, some baseline noise features observed in the measured EMI are not present in the simulated model. The measured data has thermal noise which has not been modelled in simulations. By modifying the parameters of this model, we can generate unique EMI signatures for other SMPS based appliances for further analysis.
% Second case in simulation section

\begin{figure}[t!]
  \vspace{-10pt}
  \hspace{20pt}
  \includegraphics[height=4.2cm]{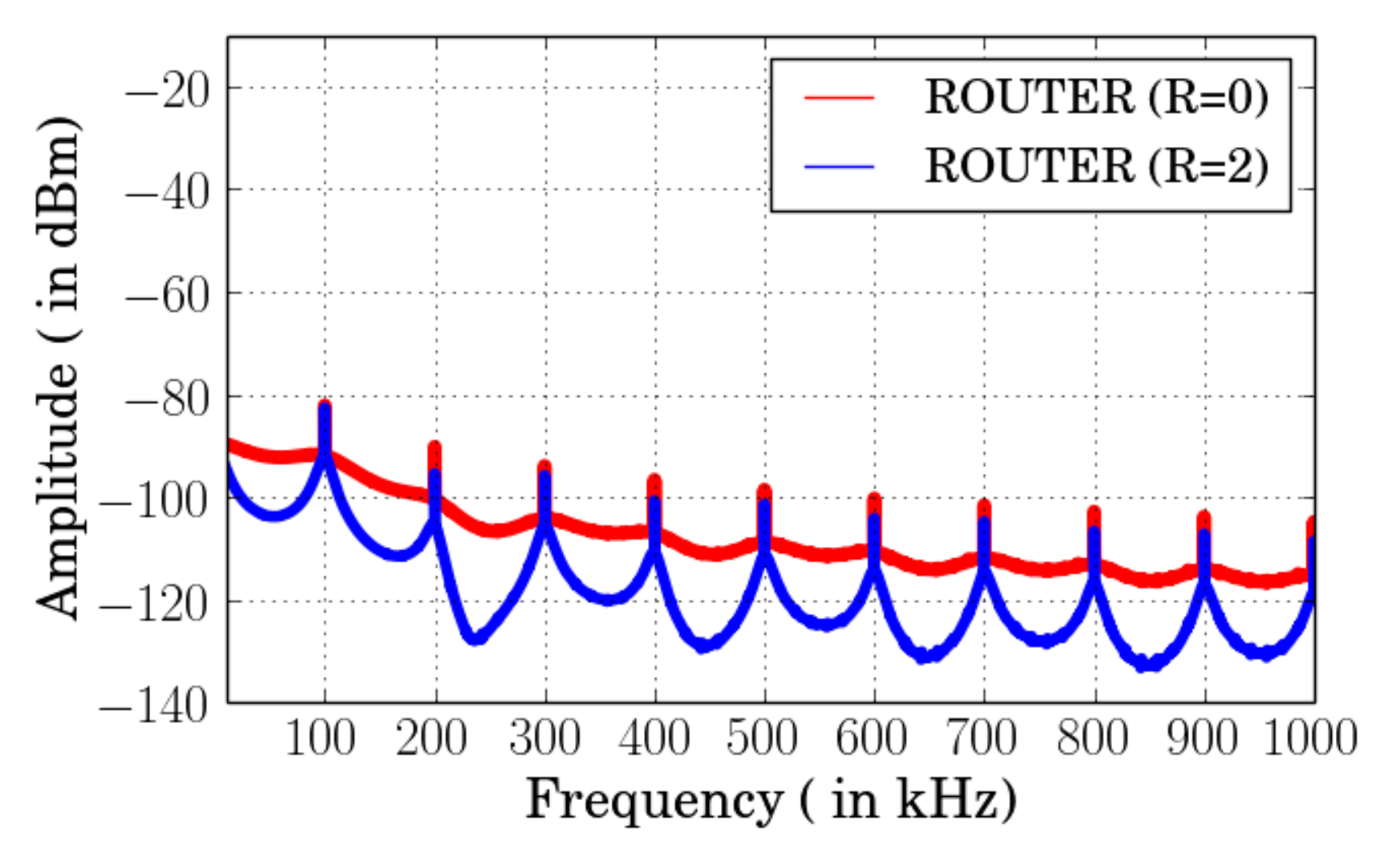}
  \vspace{-10pt}
  \caption{Effect of line impedance of R=0 \& 2$\Omega$ on EMI conducted by an appliance.}
  \vspace{-10pt}
  \label{fig:lineimpedance}
\end{figure}

\subsection{Impact of Line Impedance on EMI from an Appliance}

e model the transmission line characteristics of the power line with a series impedance between the sensing point where the EMI currents are observed (nearly the DC supply) and the plug point where the appliance is connected as shown in Figure\ref{fig:Model}. Though, the series impedance lies on the 230V AC power line in real world settings, we have introduced the equivalent resistance in the internal circuitry of the SMPS.  Due to the low frequencies of observation, the effect of the inductance and capacitance of the transmission line are negligible.  Most residential electrical lines are characterized by line impedance that varies with the thickness of wires, the number of copper strands, and the quality of shielding material \cite{russell2000impact}. We model, a 10m long electrical line with a 2$\Omega$ resistor. We compare the EMI signatures when an appliance is connected near the sensing point (when the line impedance is 0$\Omega$) with the EMI signatures generated when an appliance is connected at a distant point, 10m from the sensing point. Figure-\ref{fig:lineimpedance} shows that the EMI signal from an appliance at a distant point is weaker (up to 7 dBm) than the EMI from an appliance near the sensing point. These results are similar to the line impedance effects observed on the CFL data in Section-5c.  
% Effect of line impedance on EMI simulated from modem
        
% Effect of appliance on EMI simulated from two appliances
\begin{figure}[t!]
  \includegraphics[ height=5.5cm]{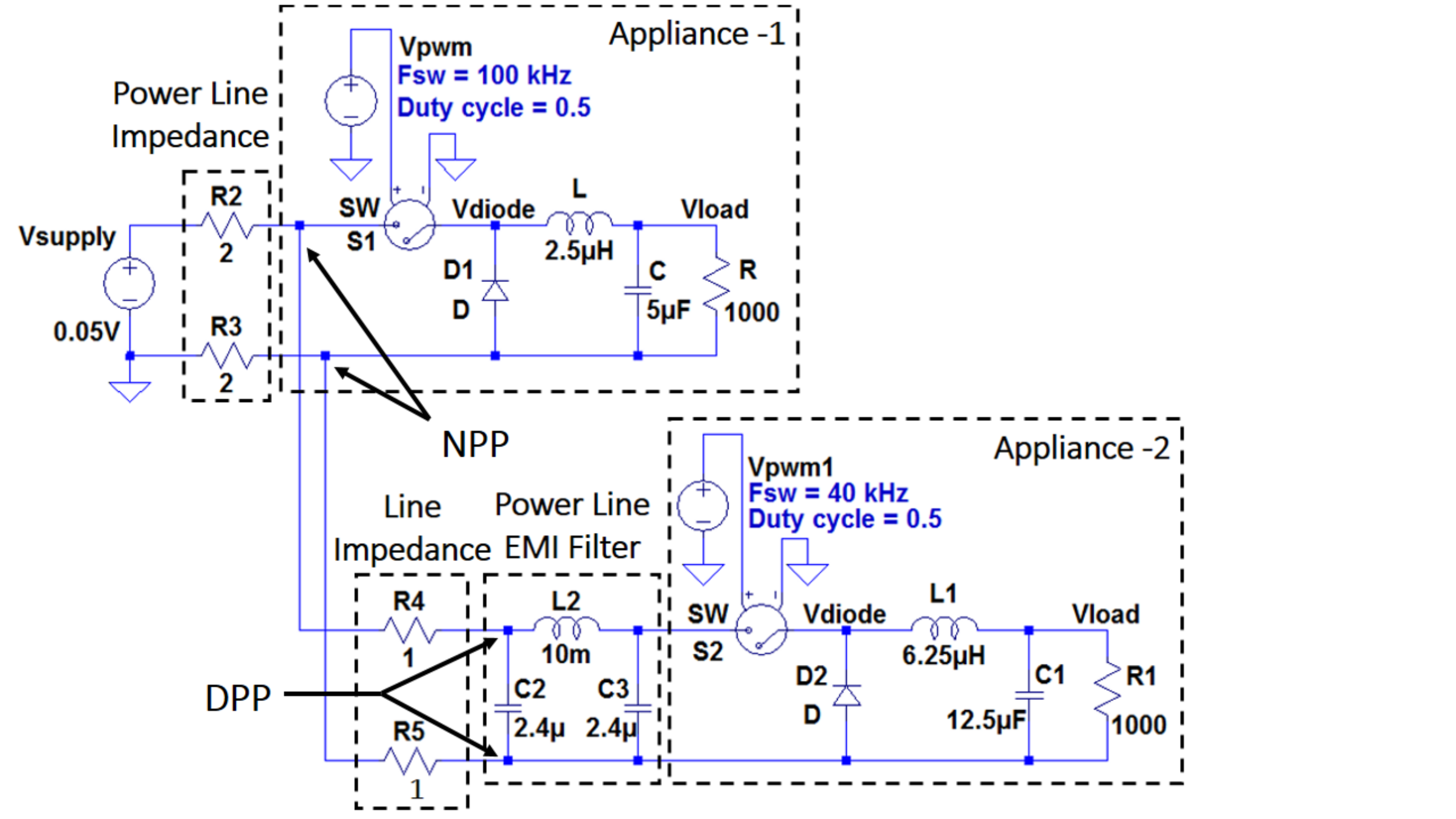}
  \vspace{-10pt}
  \caption{Model showing appliance coupling including, mains power line impedance (in Case-1, 2, 3), EMI filter (in Case-2) and line impedance \& EMI filter (in Case-3).}
  \vspace{-10pt}
  \label{fig:coupling}
\end{figure}

% Frequency spectrum showing appliance coupling
\begin{figure*}[t!]
    %\hspace{-5pt}
    \begin{subfigure}{0.375\textwidth}
    \includegraphics[width=\textwidth]{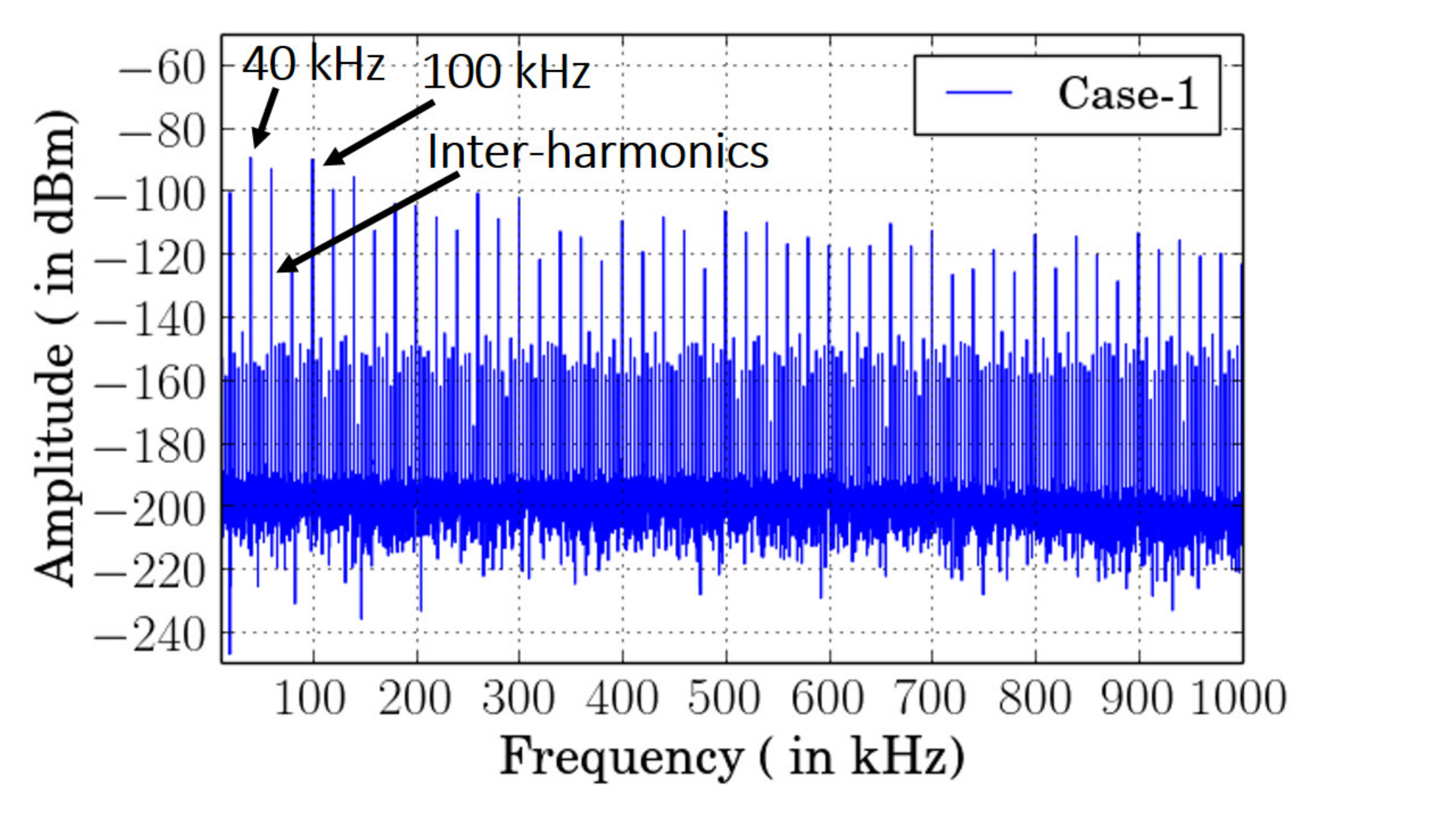}
    \vspace{-10pt}
    \caption{}
    \end{subfigure}\hspace{-20pt}
    \begin{subfigure}{0.375\textwidth}
    \includegraphics[width=\textwidth]{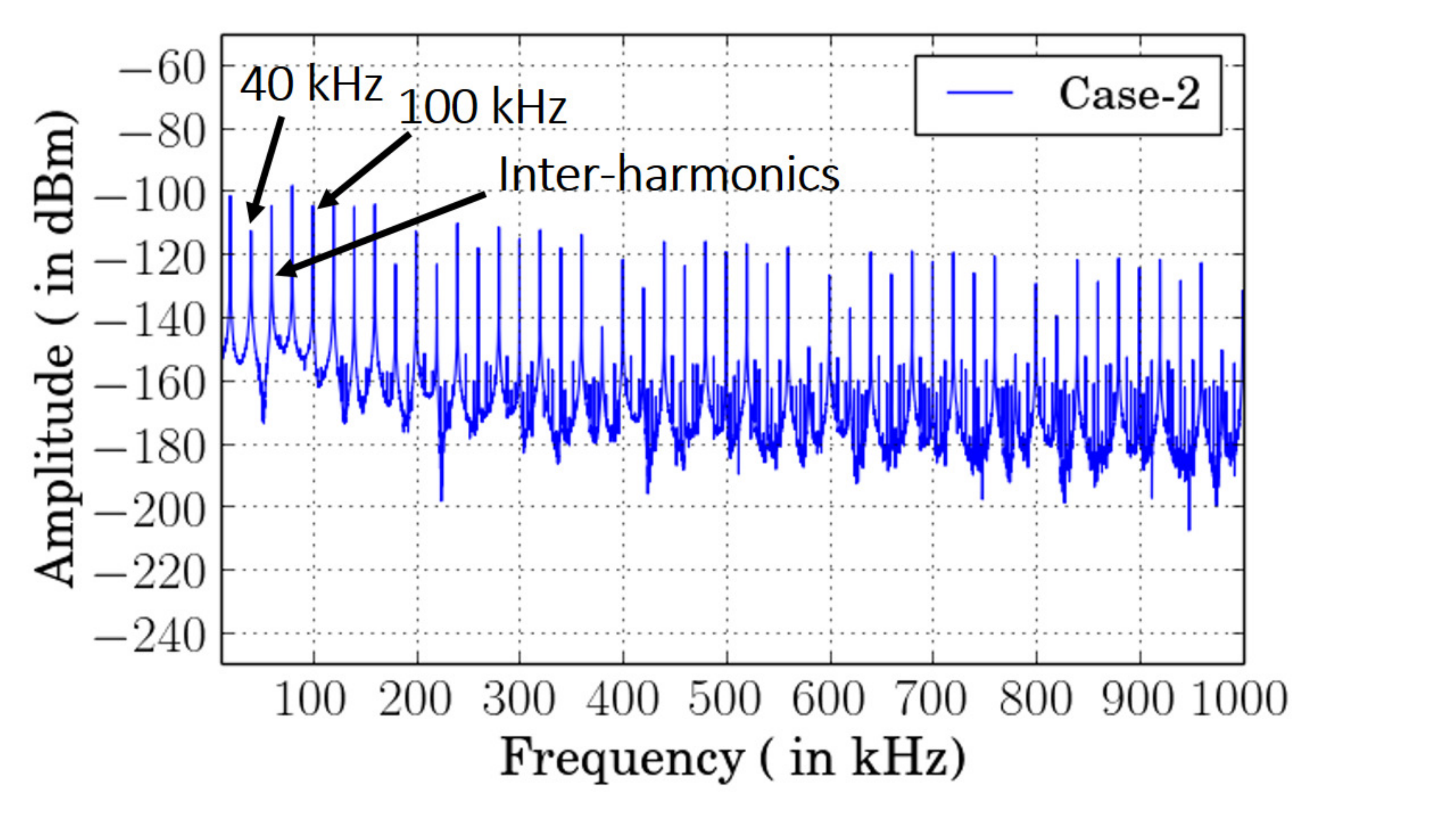}
    \vspace{-10pt}
    \caption{}
    \end{subfigure}\hspace{-20pt}
    \begin{subfigure}{0.375\textwidth}
    \includegraphics[width=\textwidth]{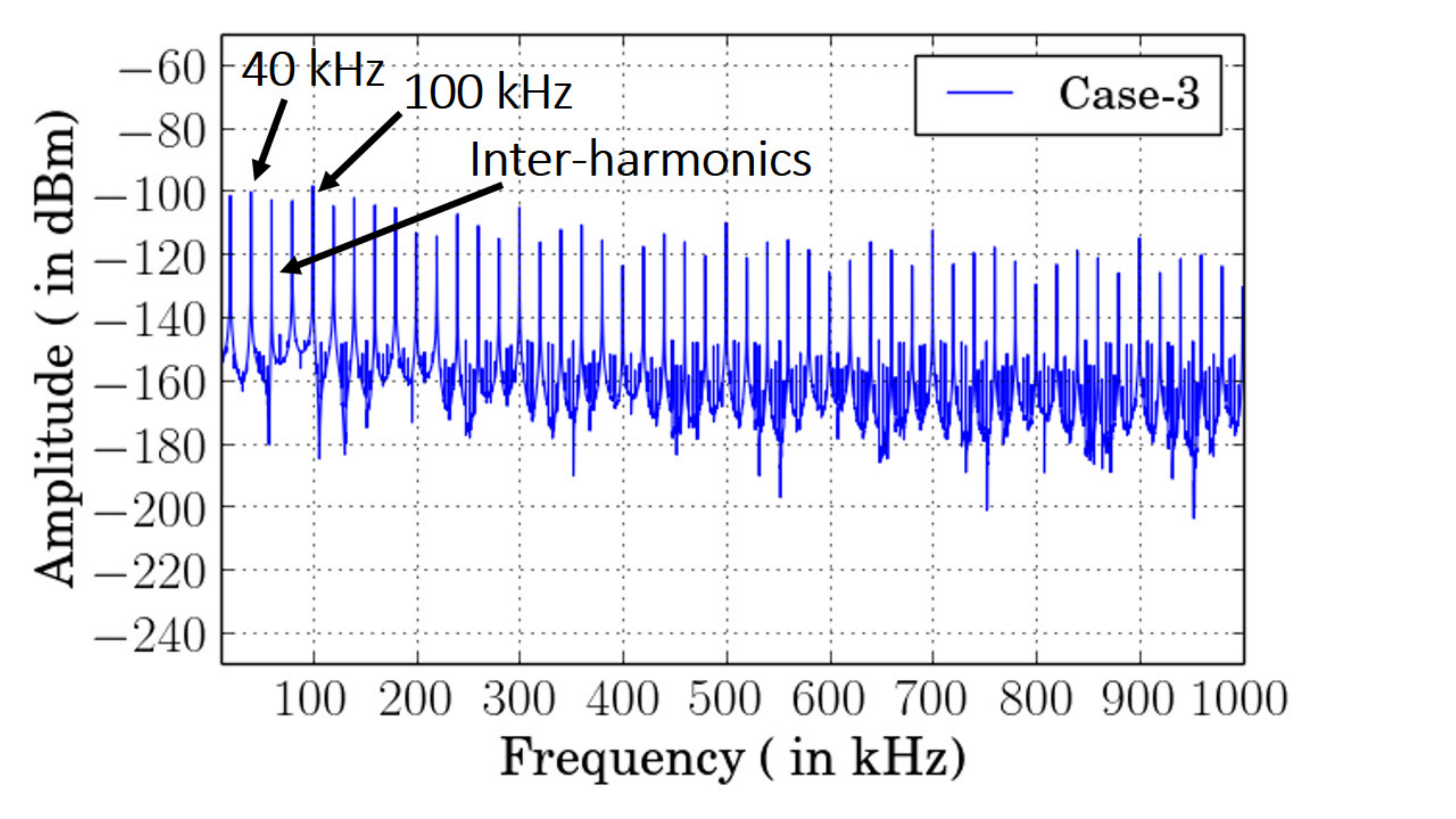}
    \vspace{-10pt}
    \caption{}
    \end{subfigure}
\vspace{-10pt}
\caption{EMI spectrum, obtained from simulation model for the three used cases - Case-1: Coupling with line impedance; Case-2: Coupling with one appliance having EMI filter;  Case-3: Coupling with line impedance and one appliance having EMI filter. }
\label{fig:FFTcoupling}
\end{figure*}                      
     
\subsection{Impact of Appliance Coupling}

In section 5.4, we postulated that an EMI filter of an appliance can significantly impact the EMI signature of a neighbouring appliance, provided the two appliances are close together on the power line. To validate this hypothesis with simulations, we consider the following three cases which are configured as per Figure-\ref{fig:coupling}. Figure-\ref{fig:FFTcoupling} shows the frequency spectrum of the EMI currents(DC supply current) for all three cases. 

\textbf{Case-1:} Two appliances-1 \& 2, without EMI filters, are connected together on the same electric line. Appliance-1 has a switching frequency (Fsw) of 100 kHz, similar to the router. Appliance-2 has Fsw of 40 kHz.   Note that in real world settings, the two appliances would be connected on the 230V AC power line. In our simulations, we instead consider the appliances connected to the same electric line from a common DC power supply. Here we are assuming that the EMI coupling mechanisms, from the DC section to AC mains power line, across different appliances are uniform. A series resistance of 4$\Omega$ is inserted between the appliances and the power supply, to model the power line impedance. Dominant frequency components in the EMI spectrum in Figure-\ref{fig:FFTcoupling} include the EMI effects from both appliances i.e. 100 kHz \& 40 kHz signals and their harmonics. The magnitudes of higher order harmonics decay gradually with frequency. Interestingly, we also observe inter-harmonics due to mixing of the 100 kHz and 40 kHz signals in the non-linear sections of the circuit (such as diode). 

\textbf{Case-2:} In this case, we consider the impact of an appliance with an inbuilt EMI filter on a neighbouring appliance. Therefore, a bi-directional EMI low pass filter, of 1 kHz cut off frequency, is incorporated into the front-end of appliance-2 (with 40 kHz switching frequency). As expected, the EMI filter suppresses the EMI currents (40 kHz and its harmonics) from appliance-2 at the sensing point in the circuit. However, the filter additionally suppresses the EMI (100 kHz and its harmonics) from appliance-1 from being observed at the sensing point. This is because the EMI filter of appliance-2 offers a low impedance path, compared to the electric line at the power supply, to the EMI currents from appliance-1. This simulation validates our earlier hypothesis, in Section-5.4, that inbuilt EMI filters within the module of an appliance have a two-fold effect on EMI based appliance disaggregation. The filters not only suppress EMI emanating from the appliance that they are connected to but also impact the measurability of EMI currents conducted by neighbouring appliances on the power line.

\textbf{Case-3:} In this case we consider how the line impedance between two appliances can impact the effect of the inbuilt EMI filter. We simulate conditions when the two appliances, considered above, are 10m apart on the power line by introducing a series resistance of 2$\Omega$ in the electric line between the two appliances. In Figure-\ref{fig:FFTcoupling}, the frequency spectrum exhibits weak EMI from appliance-2 and significant EMI (100 kHz and harmonics) from appliance-1. We observe that the impact of the EMI filter in appliance-2 on appliance-1 is reduced since the EMI currents from appliance-1 do not find a low-impedance path through the EMI filter when significant line impedance is introduced between the two appliances. 

\section{HFED Dataset and Spice Model Simulations}
    \label{sec:dataset}

Our High Frequency EMI Dataset (HFED)\footnote{\url{http://hfed.github.io}} is also released as part of this work. It contains EMI traces collected from both lab and home settings, having 24 appliances connected over four different test setups in lab settings and one test setup in home settings. EMI data from lab consists of data from signal analyser as well as USRP, while home data contains data from USRP only. EMI measurements from signal analyser and USRP are taken over a frequency range of 10 kHz to 5 MHz. 

To further facilitate reproducible research, we also release our CAD layout files and our SPICE simulations on the dataset webpage. Our dataset is summarized in Table \ref{table:HFED}.

\begin{table}
    \begin{tabular}{|l|l|}
    \hline
    Parameter & Value \\
    \hline
    Frequency range               & 10 kHz to 5 Mhz  \\
    Number of appliances tested   & 24               \\
    Number of test setups         & 5 (4 in home, 1 in lab)\\
    Number of FFT points          & 32k (Signal Analyser),\\
     ~                            & 100k (USRP)\\ \hline
    \end{tabular}
    \caption{Summary of HFED dataset}
    \label{table:HFED}
\end{table}

\section{Conclusions}
    \label{sec:conclusion}

In this paper we highlight some key observations obtained from our in-depth study of EMI signatures over an extensive set of home appliances, tested in both laboratory and residential settings. Some of these observations either counter or qualify the observations made in earlier EMI related research (e.g. not all SMPS based appliances will conduct EMI due to high attenuation offered by inbuilt EMI filters) or provide novel insights (e.g. EMI from neighbouring appliances may significantly interfere through the EMI filters and line impedance along the power line). Empirical analysis, discussed in this paper, calls for a detailed review of different factors that may impact the observed EMI from appliances before using the EMI as a signature for appliance disaggregation.  
Setting up controlled environments for better understanding of EMI sensing may either neglect real conditions - by isolating background noise and interference from other appliances on the power line; or would require a very extensive study accounting for all of these noise factors. To simplify the overall review process, accurate simulation models that account for several of these noise factors will significantly help in improved and thorough understanding. We made a first attempt in creating such simulation models and demonstrated their utility in explaining some of the peculiar behavior discussed in this paper through physics. Such simulation models can further help in analyzing minute events which are otherwise not observable with standard measuring equipment.
We believe that the domain of conducted EMI sensing is wide open and this work has only touched the surface highlighting some of the concerns brought forth by our early, though detailed, observations. 

% Future work 
\section{Future Work}
    \label{sec:Future_work}
    
While we presented results from an initial controlled experimental setup in this paper, we would like to explore EMI signatures in an in-situ setting, which we deem to be the real test of the usability of our system. Our work had been done in New Delhi, India, which is known to be affected by an unreliable electrical grid~\cite{Batra:2013:DIH:2528282.2528293}. In the future, we would like to compare our results in a setting where the electrical grid is more reliable. We would also like to study different modes in which conducted EMI can be observed - differential and common mode.
%In this work, we focused on obtaining insights from EMI measurements, based on our visual observations. Currently, we can’t make any claim of, whether a complex machine learning algorithm can still disaggregate appliances. In order to facilitate further research on conducted EMI, we have released our dataset and our code for visualization.\\ 
%Please note that this study has been conducted in New Delhi, India which suffers from unreliable grid and voltage fluctuations

%ACKNOWLEDGMENTS are optional
\section{Acknowledgments}
    \label{sec:ack}

We would like to thank Department of Electronics and Information Technology (Government of India) for funding the project (Grant Number DeitY/R\&D/ITEA/4(2)/2012 and Grant Number ITRA/15(57)/Mobile/HumanSense/01). We would also like to thank Geetali Tyagi, Raghav Sehgal and Puneet Jain (IIIT Delhi), for their support in the experimental setup, collecting EMI traces and releasing HFED dataset.


\balance
\raggedbottom
\bibliographystyle{abbrv}
\bibliography{buildsys-proc}  % buildsys-proc.bib is the name of the Bibliography in this case

\begin{thebibliography}{10}

\bibitem{TechnicalreportUK}
{\em Energy Consumption in the United Kingdom}.
\newblock Department of Energy and Climate Change, 2010.

\bibitem{barker2013empirical}
S.~Barker, S.~Kalra, D.~Irwin, and P.~Shenoy.
\newblock Empirical characterization and modeling of electrical loads in smart
  homes.
\newblock In {\em Green Computing Conference (IGCC), 2013 International}, pages
  1--10. IEEE, 2013.

\bibitem{batra2013indic}
N.~Batra, H.~Dutta, and A.~Singh.
\newblock Indic: Improved non-intrusive load monitoring using load division and
  calibration.
\newblock In {\em 12th International Conference on Machine Learning and
  Applications (ICMLA)}, volume~1, pages 79--84. IEEE, 2013.

\bibitem{Batra:2013:DIH:2528282.2528293}
N.~Batra, M.~Gulati, A.~Singh, and M.~B. Srivastava.
\newblock It's different: Insights into home energy consumption in india.
\newblock In {\em Proceedings of the 5th ACM Workshop on Embedded Systems For
  Energy-Efficient Buildings}, BuildSys'13, pages 3:1--3:8, New York, NY, USA,
  2013. ACM.

\bibitem{NILMTK}
N.~Batra, J.~Kelly, O.~Parson, H.~Dutta, W.~Knottenbelt, A.~Rogers, A.~Singh,
  and M.~Srivastava.
\newblock {NILMTK: An Open Source Toolkit for Non-intrusive Load Monitoring}.
\newblock In {\em Fifth International Conference on Future Energy Systems (ACM
  e-Energy)}, Cambridge, UK, 2014.

\bibitem{boost2007switch}
E.~Boost and R.~Dampled.
\newblock Switch-mode power supplies spice simulations and practical designs.
\newblock 2007.

\bibitem{darby2006effectiveness}
S.~Darby.
\newblock The effectiveness of feedback on energy consumption.
\newblock {\em A Review for DEFRA of the Literature on Metering, Billing and
  direct Displays}, 486:2006, 2006.

\bibitem{dong2014fundamental}
R.~Dong, L.~Ratliff, H.~Ohlsson, and S.~S. Sastry.
\newblock Fundamental limits of nonintrusive load monitoring.
\newblock In {\em Proceedings of the 3rd international conference on High
  confidence networked systems}, pages 11--18. ACM, 2014.

\bibitem{enev2011televisions}
M.~Enev, S.~Gupta, T.~Kohno, and S.~N. Patel.
\newblock Televisions, video privacy, and powerline electromagnetic
  interference.
\newblock In {\em Proceedings of the 18th ACM conference on Computer and
  communications security}, pages 537--550. ACM, 2011.

\bibitem{evans2009country}
M.~Evans, B.~Shui, and S.~Somasundaram.
\newblock {\em Country report on building energy codes in india}.
\newblock Pacific Northwest National Laboratory, 2009.

\bibitem{farhangi2010path}
H.~Farhangi.
\newblock The path of the smart grid.
\newblock {\em Power and Energy Magazine, IEEE}, 8(1):18--28, 2010.

\bibitem{froehlich2011disaggregated}
J.~Froehlich, E.~Larson, S.~Gupta, G.~Cohn, M.~Reynolds, and S.~Patel.
\newblock Disaggregated end-use energy sensing for the smart grid.
\newblock {\em IEEE Pervasive Computing}, 10(1):28--39, 2011.

\bibitem{gupta2010electrisense}
S.~Gupta, M.~S. Reynolds, and S.~N. Patel.
\newblock Electrisense: single-point sensing using emi for electrical event
  detection and classification in the home.
\newblock In {\em Proceedings of the 12th ACM international conference on
  Ubiquitous computing}, pages 139--148. ACM, 2010.

\bibitem{halverson2009country}
M.~A. Halverson, B.~Shui, and M.~Evans.
\newblock Country report on building energy codes in the united states.
\newblock {\em Pacific Northwest National Laboratory}, 2009.

\bibitem{hart1992nonintrusive}
G.~W. Hart.
\newblock Nonintrusive appliance load monitoring.
\newblock {\em Proceedings of the IEEE}, 80(12):1870--1891, 1992.

\bibitem{Survey:Mohit}
M.~Jain, D.~Chabra, J.~Mankoff, and A.~Singh.
\newblock {Energy Usage Attitudes of Urban India}.
\newblock In {\em Second International Conference ICT for Sustainability},
  Stockholm, Sweden, 2014.

\bibitem{keshav2011internet}
S.~Keshav and C.~Rosenberg.
\newblock How internet concepts and technologies can help green and smarten the
  electrical grid.
\newblock {\em ACM SIGCOMM Computer Communication Review}, 41(1):109--114,
  2011.

\bibitem{kim2010granger}
Y.~Kim, R.~Balani, H.~Zhao, and M.~B. Srivastava.
\newblock Granger causality analysis on ip traffic and circuit-level energy
  monitoring.
\newblock In {\em Proceedings of the 2nd ACM Workshop on Embedded Sensing
  Systems for Energy-Efficiency in Building}, pages 43--48. ACM, 2010.

\bibitem{kim2009viridiscope}
Y.~Kim, T.~Schmid, Z.~M. Charbiwala, and M.~B. Srivastava.
\newblock Viridiscope: design and implementation of a fine grained power
  monitoring system for homes.
\newblock In {\em Proceedings of the 11th international conference on
  Ubiquitous computing}, pages 245--254. ACM, 2009.

\bibitem{lee2004exploration}
W.~Lee, G.~Fung, H.~Lam, F.~Chan, and M.~Lucente.
\newblock Exploration on load signatures.
\newblock In {\em International conference on electrical Engineering (ICEE)},
  pages 1--5, 2004.

\bibitem{leeb1995transient}
S.~B. Leeb, S.~R. Shaw, and J.~L. Kirtley~Jr.
\newblock Transient event detection in spectral envelope estimates for
  nonintrusive load monitoring.
\newblock {\em Power Delivery, IEEE Transactions on}, 10(3):1200--1210, 1995.

\bibitem{orji2010fault}
U.~A. Orji, Z.~Remscrim, C.~Laughman, S.~B. Leeb, W.~Wichakool, C.~Schantz,
  R.~Cox, J.~Paris, J.~L. Kirtley, and L.~K. Norford.
\newblock Fault detection and diagnostics for non-intrusive monitoring using
  motor harmonics.
\newblock In {\em Applied Power Electronics Conference and Exposition (APEC),
  2010 Twenty-Fifth Annual IEEE}, pages 1547--1554. IEEE, 2010.

\bibitem{parson2012non}
O.~Parson, S.~Ghosh, M.~Weal, and A.~Rogers.
\newblock Non-intrusive load monitoring using prior models of general appliance
  types.
\newblock In {\em AAAI}, 2012.

\bibitem{patel2007flick}
S.~N. Patel, T.~Robertson, J.~A. Kientz, M.~S. Reynolds, and G.~D. Abowd.
\newblock {\em At the flick of a switch: Detecting and classifying unique
  electrical events on the residential power line (nominated for the best paper
  award)}.
\newblock Springer, 2007.

\bibitem{paul2006introduction}
C.~R. Paul.
\newblock {\em Introduction to electromagnetic compatibility}, volume 184.
\newblock John Wiley \& Sons, 2006.

\bibitem{rowe2010contactless}
A.~Rowe, M.~Berges, and R.~Rajkumar.
\newblock Contactless sensing of appliance state transitions through variations
  in electromagnetic fields.
\newblock In {\em Proceedings of the 2nd ACM Workshop on Embedded Sensing
  Systems for Energy-Efficiency in Building}, pages 19--24. ACM, 2010.

\bibitem{russell2000impact}
M.~J. Russell.
\newblock The impact of mains impedance on power quality.
\newblock In {\em Power Quality}, 2000.

\bibitem{shaw2008nonintrusive}
S.~R. Shaw, S.~B. Leeb, L.~K. Norford, and R.~W. Cox.
\newblock Nonintrusive load monitoring and diagnostics in power systems.
\newblock {\em Instrumentation and Measurement, IEEE Transactions on},
  57(7):1445--1454, 2008.

\bibitem{srinivasan2013fixturefinder}
V.~Srinivasan, J.~Stankovic, and K.~Whitehouse.
\newblock Fixturefinder: Discovering the existence of electrical and water
  fixtures.
\newblock In {\em Proceedings of the 12th international conference on
  Information processing in sensor networks}, pages 115--128. ACM, 2013.

\bibitem{taysi2010tinyears}
Z.~C. Taysi, M.~A. Guvensan, and T.~Melodia.
\newblock Tinyears: spying on house appliances with audio sensor nodes.
\newblock In {\em Proceedings of the 2nd ACM Workshop on Embedded Sensing
  Systems for Energy-Efficiency in Building}, pages 31--36. ACM, 2010.

\bibitem{zoha2012non}
A.~Zoha, A.~Gluhak, M.~A. Imran, and S.~Rajasegarar.
\newblock Non-intrusive load monitoring approaches for disaggregated energy
  sensing: A survey.
\newblock {\em Sensors}, 12(12):16838--16866, 2012.

\bibitem{zoha2013low}
A.~Zoha, A.~Gluhak, M.~Nati, and M.~A. Imran.
\newblock Low-power appliance monitoring using factorial hidden markov models.
\newblock In {\em Intelligent Sensors, Sensor Networks and Information
  Processing, 2013 IEEE Eighth International Conference on}, pages 527--532.
  IEEE, 2013.

\end{thebibliography}
\end{document}